# Stability of stratified two-phase channel flows of Newtonian/non-Newtonian shear-thinning fluids

D. Picchi[a)], I. Barmak[b)], A. Ullmann, and N. Brauner

*School of Mechanical Engineering, Tel Aviv University, Tel Aviv 69978, Israel*

Linear stability of horizontal and inclined stratified channel flows of Newtonian/non-Newtonian shear-thinning fluids is investigated with respect to all wavelength perturbations. The Carreau model has been chosen for the modeling of the rheology of a shear-thinning fluid, owing to its capability to describe properly the constant viscosity limits (Newtonian behavior) at low and high shear rates. The results are presented in the form of stability boundaries on flow pattern maps (with the phases' superficial velocities as coordinates) for several practically important gas-liquid and liquid-liquid systems. The stability maps are accompanied by spatial profiles of the critical perturbations, along with the distributions of the effective and tangent viscosities in the non-Newtonian layer, to show the influence of the complex rheological behavior of shear-thinning liquids on the mechanisms responsible for triggering instability. Due to the complexity of the considered problem, a working methodology is proposed to alleviate the search for the stability boundary. Implementation of the proposed methodology helps to reveal that in many cases the investigation of the simpler Newtonian problem is sufficient for the prediction of the exact (non-Newtonian) stability boundary of smooth stratified flow (i.e., in case of horizontal gas-liquid flow). Therefore, the knowledge gained from the stability analysis of Newtonian fluids is applicable to those (usually highly viscous) non-Newtonian systems. Since the stability of stratified flow involving highly viscous Newtonian liquids has not been researched in the literature, interesting findings on the viscosity effects are also obtained. The results highlight the limitations of applying the simpler and widely used power-law model for characterizing the shear-thinning behavior of the liquid. That model would predict a rigid layer (infinite viscosity) at the interface, where the shear rates in the viscous liquid are low, and thereby unphysical representation of the interaction between the phases.

## I. INTRODUCTION

Stratified flows of Newtonian/non-Newtonian fluids can be found in various industrial applications, such as chemical industry, petroleum transport in pipelines and polymer extrusion. The stability analysis is essential tool for the prediction of operating conditions for which stratified flow with a smooth interface is stable. In fact, instability may result in transition from stratified-smooth to stratified-wavy flow or to other flow patterns (e.g., slug flow, annular flow). However, only a few studies on the stability of two-phase flows where one of the phases is a non-Newtonian liquid (e.g., shear-thinning liquid) are available in the literature, and this issue has not been sufficiently investigated yet.

---

[a)] Present address: Energy Resources Engineering, Stanford University, Stanford, CA 94305, USA.

[b)] Author to whom correspondence should be addressed. Electronic mail: ilyab@tauex.tau.ac.il (I. Barmak).

Several liquids of practical importance (e.g. waxy oils, dense emulsions and polymer solutions) exhibit a non-Newtonian shear-thinning behavior. For this type of liquids, the effective viscosity is a function of the imposed shear rate. The Carreau (1972) viscosity model is considered as an appropriate viscosity model to describe the rheology of the non-Newtonian liquid due to its capability to describe the zero-shear-rate and the infinity-shear-rate Newtonian viscosity limits. As recently shown by Picchi et al (2017), considering the idealized Ostwald – de Waele power-law viscosity model for predicting the integral flow characteristics (e.g., holdup, pressure drop) can lead to erroneous results, since this rheological model predicts unlimited growth of the effective viscosity at low shear rates. In fact, for such operating conditions the non–Newtonian liquid behaves practically as a Newtonian liquid.

In spite of its importance for industrial applications, the effect of the rheology of a shear-thinning fluid on the characteristics of the stability of two-phase stratified flow is still not sufficiently elaborated as only few publications addressed this topic. Since the exact formulation of transient flow in pipes is too complicated for conducting a rigorous stability analysis, a common approach is to use the transient one-dimensional Two-Fluid (TF) mechanistic model (Picchi et al., 2014; Picchi and Poesio, 2016b). However, the predictions obtained via the Two-Fluid model critically depend on the reliability of the closure relations used to model the shear stresses (see Picchi and Poesio, 2016a), and the results are valid only under the long-wave assumption. The non-Newtonian rheology of the liquids introduces additional complexity that further deteriorates the TF model predictions.

An alternative approach for obtaining an insight into the mechanisms involved in the flow destabilization is a rigorous stability analysis of stratified flow in the Two Plates (TP) geometry, while considering all wavenumber perturbations. This approach was widely used for Newtonian fluids (e.g., see the pioneer works of Yih, 1967; Yiantsios and Higgins, 1988). More recently, Barmak et al. (2016a, b) conducted a comprehensive stability analysis of horizontal and inclined stratified two-phase flows in the TP geometry providing stability boundaries on the flow pattern maps and physical interpretations of the flow pattern transition associated with the flow destabilization. The TP geometry was used also to study the stability of stratified flow involving a non-Newtonian liquid. However, except for Pinarbasi and Liakopoulos (1995), who considered the case of two-Carreau fluids, the majority of the published works on the stability of stratified flow with a shear-thinning liquid addressed only the idealized power-law or Herschel-Bulkley viscosity models (e.g., Khomami, 1990; Su and Khomami, 1991; Sahu el al., 2007; Ó Náraigh and Spelt, 2010; Alba et al., 2013).

In this study, a linear stability analysis of stratified flow of Newtonian/non-Newtonian shear-thinning fluids with respect to all wavelength perturbations is carried out for horizontal and inclined channels. The shear-thinning fluid is modeled as a Carreau fluid. Our main scope is to obtain a physical insight on the instability mechanisms that may trigger flow pattern transitions. To the best of our knowledge, such an analysis has not been carried out yet.

The results are presented in the form of stability boundaries on flow pattern maps referring to some practically important cases for both gas-liquid and liquid-liquid systems. Along with the profiles of the critical perturbation that are responsible for triggering the instability, the cross-section distribution of the effective viscosity is demonstrated to offer a better physical insight on the effect of rheology on the predictions. In addition, a working methodology is proposed to simplify the search for the stability boundary, which in some cases (i.e., in case of horizontal gas-liquid flow) can be predicted by considering the simpler Newtonian fluids problem. We also discuss the applicability of using a stability criterion that is based on the Two-fluid model (Kushnir et al., 2017) for the prediction of the stratified-smooth boundary for the studied two-phase systems.

## II. PROBLEM FORMULATION

We consider a stratified two-layer channel flow of two immiscible incompressible fluids. The flow configuration in an inclined channel $(0 \leq \beta < \pi/2)$ is shown schematically in Fig. 1. The flow, assumed isothermal and two-dimensional, is driven by an imposed pressure gradient and a component of the gravitational force in the *x*-direction. One of the phases is a shear-thinning liquid of negligible visco-elasticity, and the other is a Newtonian fluid. The interface between the fluids, labeled as $j=1,2$ (1 – non-Newtonian phase, 2 – Newtonian phase), is assumed to be flat in the undisturbed base-flow state. Under this assumption, the model allows for a plane-parallel solution, in which position of the interface is an unknown value (to be determined below).

The flow field in each of the fluids is described by the continuity and momentum equations that are rendered dimensionless, choosing for the scales of length and velocity the height of the Newtonian layer $h_2$ and the interfacial velocity $u_i$, respectively. The time and the pressure are scaled by $h_2/u_i$, and $\rho_2 u_i^2$, respectively. The dimensionless continuity and momentum equations are

$$\nabla \cdot \mathbf{u}_j = 0,$$
$$\frac{\partial \mathbf{u}_1}{\partial t} + (\mathbf{u}_1 \cdot \nabla)\mathbf{u}_1 = -\frac{1}{r}\nabla p_1 + \frac{m}{\text{Re}_2 \, r}\nabla \cdot \boldsymbol{\tau}_1 + \frac{\sin\beta}{\text{Fr}_2}\mathbf{e}_x - \frac{\cos\beta}{\text{Fr}_2}\mathbf{e}_y, \qquad (1)$$
$$\frac{\partial \mathbf{u}_2}{\partial t} + (\mathbf{u}_2 \cdot \nabla)\mathbf{u}_2 = -\nabla p_2 + \frac{1}{\text{Re}_2}\Delta \mathbf{u}_2 + \frac{\sin\beta}{\text{Fr}_2}\mathbf{e}_x - \frac{\cos\beta}{\text{Fr}_2}\mathbf{e}_y,$$

where $\mathbf{u}_j = (u_j, v_j)$ and $p_j$ are the velocity and pressure of the fluid $j$, $\rho_j$ and $\mu_j$ are the corresponding density and dynamic viscosity; $\mathbf{e}_x$ and $\mathbf{e}_y$ are the unit vectors in the direction of the x- and y-axes.

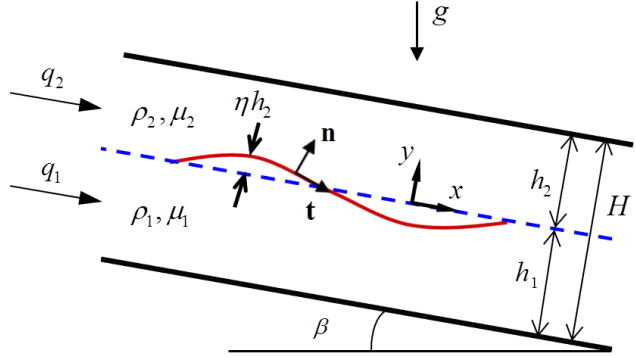

FIG. 1. Configuration of stratified two-layer channel flow.

In the dimensionless formulation the lower and upper phases occupy the regions $-n_h \leq y \leq 0$, and $0 \leq y \leq 1$, respectively, $n_h = h_1/h_2$, and $h = h_1/H$ (i.e., the holdup of the non-Newtonian phase). Other dimensionless parameters: $Re_2 = \rho_2 u_i h_2 / \mu_2$ is the Reynolds number of the Newtonian phase, $Fr_2 = u_i^2 / g h_2$ is the Froude number, $r = \rho_1/\rho_2$ and $m = \mu_0/\mu_2$ are the density and (zero shear rate) viscosity ratios.

It is important to note that for gas-liquid systems the lighter Newtonian phase is on top of the denser non-Newtonian phase (the gravitational force points from phase "2" to "1", $g > 0$). In liquid-liquid (e.g., oil-water) systems, however, the non-Newtonian fluid is typically the lighter oil phase. For the analysis of such cases the gravitational force points from phase "1" to "2" (i.e., $g < 0$), while $r = \rho_1/\rho_2 < 1$. With this formulation, the Rayleigh-Taylor instability is not encountered.

In order to model the rheological behavior of a shear-thinning liquid we have chosen the Carreau model, owing to its capability to describe properly the constant viscosity limits at low shear rates (zero-shear-rate viscosity, $\mu_0$) and high shear rates (infinity-shear-rate viscosity $\mu_\infty$). In addition, the Carreau model gives a good fit for many polymer solutions used in applications (e.g., Carboxymethyl cellulose).

The constitutive relation of a Carreau liquid is

$$\boldsymbol{\tau}_1 = \mu \dot{\boldsymbol{\gamma}} \quad \text{with} \quad \mu = \frac{\mu_1}{\mu_0} = \frac{\mu_\infty}{\mu_0} + \left[1 - \frac{\mu_\infty}{\mu_0}\right]\left[1 + \left(\tilde{\lambda}_C \dot{\gamma}\right)^2\right]^{(n-1)/2}, \tag{2}$$

where $\boldsymbol{\tau}$ is the stress tensor and $\dot{\boldsymbol{\gamma}}$ is the strain-rate tensor, that has the following components

$$\dot{\gamma}_{kl}(\mathbf{u}) = \frac{\partial u_k}{\partial x_l} + \frac{\partial u_l}{\partial x_k}, \tag{3}$$

and the norm of the $\dot{\boldsymbol{\gamma}}$ tensor is defined by

$$\dot{\gamma}(\mathbf{u}) = \left[\frac{1}{2}\sum_{k,l=1}^{2}\left[\dot{\gamma}_{kl}(\mathbf{u})\right]^2\right]^{1/2}. \tag{4}$$

In Eq. (2) $\mu_\infty$ is the infinite-shear-rate viscosity and $\mu_0$ is the zero-shear-rate viscosity (i.e., liquid behaves as a Newtonian at very low and very high shear rates), while $\boldsymbol{\tau}_1$ is the dimensionless shear stress scaled by $\mu_0 u_i / h_2$. The power law index $n$ represents the degree of shear-thinning (for a Newtonian liquid $n = 1$ and $\mu_1 = \mu_0$), whereas its onset depends on the time constant of the liquid $\lambda_C$, or in the dimensionless form $\tilde{\lambda}_C = \lambda_C \cdot u_i / h_2$ (see Eq. (2)). For higher values of $\lambda_C$ the onset of the shear-thinning behaviour shifts to lower shear rates and vice versa.

The velocities satisfy the no-slip boundary conditions at the channel walls

$$\mathbf{u}_1(y = -n_h) = 0, \quad \mathbf{u}_2(y = 1) = 0. \tag{5}$$

Boundary conditions at the interface $y = \eta(x,t)$ require continuity of velocity components and the tangential stresses, and a jump of the normal stress due to the surface tension (the jump of the quantity $f$ across the interface is denoted by $[f] = f_2 - f_1$)

$$\mathbf{u}_1(y = 0) = \mathbf{u}_2(y = 0), \tag{6}$$

$$[\mathbf{t} \cdot \boldsymbol{\tau} \cdot \mathbf{n}] = \left[\frac{m\mu^*}{\mu_0}\left\{\left(\frac{\partial u}{\partial y}+\frac{\partial v}{\partial x}\right)\left(1-\left(\frac{\partial \eta}{\partial x}\right)^2\right)-4\frac{\partial u}{\partial x}\frac{\partial \eta}{\partial x}\right\}\right] = 0, \quad (7)$$

$$[\mathbf{n} \cdot \boldsymbol{\tau} \cdot \mathbf{n}] = \left[p + \frac{m\mu^*}{\mu_0}\frac{2\operatorname{Re}_2^{-1}}{1+\left(\frac{\partial \eta}{\partial x}\right)^2}\left(\frac{\partial u}{\partial x}\left(1-\left(\frac{\partial \eta}{\partial x}\right)^2\right)+\left(\frac{\partial u}{\partial y}+\frac{\partial v}{\partial x}\right)\frac{\partial \eta}{\partial x}\right)\right]$$

$$= \operatorname{We}_2^{-1}\frac{\frac{\partial^2 \eta}{\partial x^2}}{\left(1+\left(\frac{\partial \eta}{\partial x}\right)^2\right)^{3/2}}, \quad (8)$$

where $\mathbf{n}$ is the unit normal vector pointing from the lower into the upper phase, $\mathbf{t}$ is the unit vector tangent to the interface; $\mu^* = \mu$ for the non-Newtonian phase and $\mu^* = \mu_2$ for the Newtonian one. The additional dimensionless parameter $\operatorname{We}_2 = \rho_2 h_2 u_i^2 / \sigma$ is the Weber number, and $\sigma$ is the surface tension coefficient.

Additionally, the interface displacement and the normal velocity components at the interface satisfy the kinematic boundary condition

$$v_j = \frac{D\eta}{Dt} = \frac{\partial \eta}{\partial t} + u_j \frac{\partial \eta}{\partial x}. \quad (9)$$

### III. BASE FLOW

The base-flow solution is obtained for laminar, steady and fully developed conditions. In this section, we follow the normalisation used in Picchi et al. (2017) (indicated by a subscript "BF"). It is assumed that $U(y_{BF})$ varies only in the $y$ direction (unidirectional flow) and can be obtained by solving the following momentum balances

$$-h \leq y_{BF} \leq 0: \quad \frac{\partial}{\partial y_{BF}}\left\{m\left[\frac{\mu_\infty}{\mu_0}+\left[1-\frac{\mu_\infty}{\mu_0}\right]\left[1+\left(\lambda_{C,BF}\left|\frac{dU_1}{dy_{BF}}\right|\right)^2\right]^{(n-1)/2}\right]\frac{dU_1}{dy_{BF}}\right\} = 12(\tilde{P}-Y) \quad (10)$$

$$0 \leq y_{BF} \leq 1-h: \quad \frac{d^2 U_2}{dy_{BF}^2} = 12\tilde{P} \quad (11)$$

where $\tilde{P} = \dfrac{dP/dx - \rho_2 g \sin\beta}{(-dP/dx)_{2S}}$ is the dimensionless pressure gradient, $Y = \dfrac{\rho_2(r-1)g\sin\beta}{(-dP/dx)_{2S}}$ is the inclination parameter, $\lambda_{C,BF} = \dfrac{\lambda_C |U_{2S}|}{H}$, $y_{BF} = \dfrac{y}{H}$, $U_{jS} = \dfrac{q_j}{H}$.

The velocity of each phase, $U_J$, is normalized by the Newtonian superficial velocity, $U_{2S}$, and $(-dP/dx)_{jS} = 12\mu_j q_j / H^3$ is the corresponding superficial pressure drop for single-phase flow in the channel, where $H = h_1 + h_2$. The volumetric flow rate, $q_j$, and the superficial velocity are used interchangeably in the following discussion. Referring to Fig. 1, in horizontal and concurrent downward inclined flow the flow rates of both phases are positive, while in case of upward inclined flow the flow rates are both negative (the inclination angle $\beta$ is always considered positive). In countercurrent flows, the heavy phase flows downward and the light phase flows upward $(q_1/q_2 < 0)$.

The problem is subjected to the flow rate constrains, which read:

$$\int_{-h}^{0} U_1(y_{BF}) dy_{BF} = q, \tag{12}$$

$$\int_{0}^{1-h} U_2(y_{BF}) dy_{BF} = 1, \tag{13}$$

where $q = q_1/q_2$ is the flow rate ratio.

The solution of the system of equations (10) and (11), coupled by the boundary conditions at the interface and the channel walls and the two flow rate constrains (Eqs. (12) and (13)), can be found in Picchi et al. (2017). The conditions for which a solution for this problem is guaranteed were discussed by Frigaard and Scherzer (1988). Note that the problem is fully determined by eight dimensionless parameters $(m, \tilde{P}, Y, q, h, n, \lambda_{C,BF}, \mu_\infty/\mu_0)$, and, in this work, the holdup and dimensionless pressure gradient are calculated in terms of all other dimensionless parameters, i.e. for a known flow rate ratio (known flow rates of the phases), geometry, and fluid properties.

An iterative scheme is applied to find the solution. The algorithm consists of an outer iteration loop to compute $h$ that satisfies the prescribed $q$ value, i.e., $\Delta q = q - q_{cal}(\tilde{P}, h) = 0$, where $\tilde{P}$ and $q_{cal}$ are obtained by two inner iteration loops to satisfy the flow rate constrains (Eqs. (12) and (13)). The pressure gradient $\tilde{P}$ is found by solving the mass conservation of the Newtonian phase, Eq. (13), in the form $F_2(\tilde{P}, h) = 0$. Then, the 'intermediate' flow rate ratio, $q_{cal}$, is computed in another loop solving the mass conservation of the non-Newtonian phase, Eq. (12), in the form $F_1(q_{cal}, \tilde{P}, h) = 0$ (the integral in Eq. (12) is calculated numerically). In these loops, the solution of an implicit algebraic equation is carried out using the Brent-Dekker method (see Brent, 1971). The values of $h$ (and $q_{cal}$) that satisfy $\Delta q = 0$ (outer loop) are the solution.

Both of the iteration loops contain an inner loop, that requires the no-slip condition at the interface $(U_{i,1} - U_{i,2} = 0)$ to find the interfacial shear stress $\tau_i$. In fact, when $(\tilde{P}, h)$ are known, the no-slip condition is a monotonic function of $\tau_i$ (Alba et al. 2013), while the interfacial velocities computed for each layer are a function of $(\tau_i, \tilde{P}, h)$. $U_{i,2}$ is obtained analytically, $U_{i,2} = 6\tilde{P}h^2 + (\tau_i - 12\tilde{P}h)h + (-6\tilde{P} - \tau_i + 12\tilde{P}h)$, while $U_{i,1}$ is calculated by solving numerically Eq. (10). The latter is discretized by the Chebyshev collocation points method (see Canuto et al., 2006), and the velocity profile is found by the Broyden's method (see Quaternoni et al., 2007) imposing the no-slip condition at the wall and the interfacial stress exerted by the Newtonian layer. In most of the cases it is sufficient to use 12 collocation points for the non-Newtonian layer to assure convergence of the steady state solution. The Broyden's method requires an initial guess, which has been chosen as a Newtonian velocity profile with the same viscosity ratio $m$ and multiplied by an arbitrary constant (to be tuned to assure convergence).

It can be easily guaranteed that the iterative scheme gives always a solution since the flow constrain loops have only a single solution for a given value of $h$. However, the outer loop can have multiple-holdup solution (for specified flow rates of the two phases) in inclined flows, while there is always only one solution for horizontal flow. In the case of $n=1$ (two Newtonian fluids), Eq. (10) can be reduced to the simplified (Newtonian) form. In this case, the numerical solution converges to the analytical solution for Newtonian stratified two-phase flow, which can be found in the literature (e.g., in Kushnir et al., 2014).

## IV. LINEAR STABILITY

In the following, we study the linear stability of the above plane-parallel solutions with respect to infinitesimal, two-dimensional disturbances. The perturbed velocities and pressure fields are written as $u_j = U_j + \tilde{u}_j$, $v_j = \tilde{v}_j$, $p_j = P_j + \tilde{p}_j$, and $\eta = \tilde{\eta}$ for the dimensionless disturbance of the interface. The shear stress perturbation for a unidirectional base flow can be expressed as (for more details see Nouar et al., 2007)

$$\tilde{\tau}_{kl} = \begin{cases} \mu(U_1)\dot{\gamma}_{kl}(\tilde{u}_1) & \text{for } kl = xx, yy, \\ \mu_t(U_1)\dot{\gamma}_{kl}(\tilde{u}_1) & \text{for } kl = xy, yx, \end{cases} \quad (14)$$

where $\mu(U_1)$ is the (dimensionless) effective viscosity (see Eq. (2)) and $\mu_t(U_1)$ is the (dimensionless) tangent viscosity given by

$$\mu_t(U_1) = \mu(U_1) + \left.\frac{d\mu}{d\dot{\gamma}_{xy}}\right|_{U_1} \dot{\gamma}_{xy}(U_1). \quad (15)$$

The tangent viscosity is defined by $\mu_t = d\tau_{xy}/d\dot{\gamma}_{xy}$ and is used to simplify the linear stability formulation (for shear-thinning liquids $\mu_t < \mu$, for Newtonian liquids $\mu_t = \mu = 1$). Note that the stress tensor perturbation $\tilde{\boldsymbol{\tau}}_1$ is anisotropic for non-linear viscous fluids (see Nouar et al., 2007).

The disturbed velocities are conveniently represented by the corresponding stream function $\left(\tilde{u}_j = \partial\psi_j/\partial y \,;\, \tilde{v}_j = -\partial\psi_j/\partial x\right)$, and an exponential dependence of the perturbation in time is assumed

$$\begin{pmatrix} \psi_j \\ \tilde{p}_j \\ \eta \end{pmatrix} = \begin{pmatrix} \phi_j(y) \\ f_j(y) \\ H_\eta \end{pmatrix} e^{(ikx+\lambda t)}, \quad \begin{pmatrix} \tilde{u}_j \\ \tilde{v}_j \end{pmatrix} = \begin{pmatrix} \phi'_j \\ -ik\phi_j \end{pmatrix} e^{(ikx+\lambda t)}, \quad (16)$$

where $\phi_j$, $f_j$ and $H_\eta$ are the perturbation amplitudes, $k$ is the dimensionless real wavenumber ($k = 2\pi h_2/l_{wave}$, with $l_{wave}$ being the wavelength) and $\lambda$ is the complex time increment.

The formulation assumes 2D disturbances. Note, however, that for non-linear viscous fluids (e.g., shear-thinning liquids) there is no equivalence for the Squire's theorem (named after Squire, 1933), which was formulated for Newtonian fluids and states the sufficiency of consideration of 2D (in the plane of flow) perturbations for stability analysis, since they are the critical perturbations. Only recently, the applicability of the Squire's theorem for inclined two-phase Newtonian systems was provided by Barmak et al. (2017). In the presence of a non-Newtonian liquid, this issue was only verified numerically (e.g., see Nouar and Frigaard, 2009; Sahu and Matar, 2010; Allouche et al., 2015).

To apply collocation spectral method based on the Chebyshev polynomials (defined in the interval $[0,1]$), a new coordinate $y_1 = (y+n_h)/n_h$ $(0 \le y_1 \le 1)$ should be introduced for the part of the channel occupied by the non-Newtonian phase, while $y_2 = y$ $(0 \le y_2 \le 1)$ for the Newtonian layer remains unchanged.

Upon substitution and linearization of the original equations and boundary conditions (1)-(6), the problem is reduced to the Orr-Sommerfeld equations, written here in the eigenvalue problem form

$0 \le y_1 \le 1$
$(-n_h \le y \le 0)$:
$$\lambda(D_1^2 - k^2)\phi_1 = ik\left[-U_1(D_1^2 - k^2) + \frac{U_1''}{n_h^2}\right]\phi_1$$
$$+ \frac{m}{\text{Re}_2 r}\left[(D_1^2\mu_t + 2D_1\mu_t D_1 + \mu_t D_1^2)(D_1^2 + k^2)\right. \quad (17)$$
$$\left. + k^2\mu_t(D^2 + k^2) - 4k^2(D\mu D + \mu D^2)\right]\phi_1,$$

$0 \le y_2 \le 1$
$(0 \le y \le 1)$:
$$\lambda(D_2^2 - k^2)\phi_2 = \left[ik(-U_2(D_2^2 - k^2) + U_2'') + \frac{1}{\text{Re}_2}(D_2^4 - 2k^2 D_2^2 + k^4)\right]\phi_2, \quad (18)$$

where $D_1\phi_1 = \dfrac{\phi_1'}{n_h}$, $D_1^2\phi_1 = \dfrac{\phi_1''}{n_h^2}$; $D_2\phi_2 = \phi_2'$, $D_2^2\phi_2 = \phi_2''$.

The linearized boundary conditions are obtained by means of Taylor expansions of $\eta$ around its unperturbed zero value

$y_1 = 1,$
$y_2 = 0$:
$$\lambda H_\eta = -ik(\phi_2 + U_2 H_\eta), \quad (19)$$

with $H_\eta = \dfrac{\phi_2'(0) - \phi_1'(1)/n_h}{U_1'(1)/n_h - U_2'(0)}$,

$y_1 = 1,$
$y_2 = 0$:
$$\lambda\left(r \cdot \frac{\phi_1'(1)}{n_h} - \phi_2'(0)\right) = ik\left[-\left(k^2 \text{We}_2^{-1} + \frac{\cos\beta}{\text{Fr}_2}(r-1)\right) \cdot H_\eta\right.$$
$$+ r\left(-U_1\frac{\phi_1'(1)}{n_h} + \frac{U_1'(1)}{n_h}\phi_1(1)\right) + (U_2\phi_2'(0) - U_2'(0)\phi_2(0))\right] \quad (20)$$
$$+ \frac{1}{\text{Re}_2}\left[m\left((D_1\mu_t + \mu_t(1)D_1)(D_1^2 + k^2)\phi_1 - 4k^2\mu(1)D\phi_1\right) - (\phi_2'''(0) - 3k^2\phi_2'(0))\right],$$

$y_1 = 0$
$(y = -n_h)$:
$$\phi_1 = \phi_1' = 0, \quad (21)$$

$y_2 = 1$:
$$\phi_2 = \phi_2' = 0, \quad (22)$$

$y_1 = 1,$
$y_2 = 0$:
$$\phi_1(1) = \phi_2(0), \quad (23)$$

$$y_1 = 1,$$
$$y_2 = 0: \quad m\left[\mu_t(1)\left(D_1^2 + k^2\right)\phi_1 + \mu_t(1)\frac{U_1''}{n_h^2}H_\eta\right] = \phi_2''(0) + k^2\phi_2(0) + U_2''H_\eta. \quad (24)$$

Note that additional terms accounting for the shear-thinning behavior of the non-Newtonian liquid (i.e., terms involving $\mu$ and $\mu_t$) appear in Eqs. (17), (20) and (24). For $\mu_t = \mu = 1$, the governing equations and boundary conditions presented above reduce to those obtained for Newtonian fluids (Barmak et al., 2016a, b).

The temporal linear stability is studied by solving the system of differential equations (17), (18) and (19)-(24) assuming an arbitrary wavenumber for each given set of the other parameters. The time increment is defined as a complex eigenvalue $\lambda = \lambda_R + i\lambda_I$, where $\lambda_R$ determines the growth rate of the perturbation. When $u_i > 0$, the flow is considered to be stable if the real parts of all eigenvalues are negative. On the other hand, the flow with $u_i < 0$ is stable only if all eigenvalues are positive, since the time is scaled by $h_2 / u_i$, which becomes negative in this case. Neutral stability corresponds to $\max(\lambda_R) = 0$ for $u_i > 0$ ($\min(\lambda_R) = 0$ for $u_i < 0$). The dimensionless phase speed of the perturbation is determined by the quantity $c_R = -\lambda_I / k$, where $\lambda_I$ is the wave angular frequency. The stable stratified flow region corresponds to conditions for which perturbations with all wavenumbers $k$ are damped. Along the stability boundary, a perturbation of one particular wavenumber is neutrally stable (i.e., the critical perturbation), while beyond that boundary there are many unstable perturbations.

The stability problem is solved by applying the Chebyshev collocation method (with $N = 50$ collocation points for each sublayer) for discretization of the Orr-Sommerfeld equations and the boundary conditions (see details in Barmak et al., 2016a) and by using the QR algorithm (Francis, 1962) for the computation of the eigenvalues and eigenvectors. The numerical solution was verified by a comparison with the solution for the flow of two Newtonian fluids (presented in Barmak et al., 2016a, b) and by assuring the numerical convergence (independency of the results on the truncation number, $N$). The latter is demonstrated in Table I for points on the stability boundary of a gas-liquid system and a liquid-liquid system (both points are in the shear-thinning region, see Figures 2 and 12 below). Note that differently from Newtonian systems, carrying out the stability analysis for shear-thinning liquids requires the calculation of the Chebyshev polynomials coefficients (see Quaternoni and Valli, 1994) corresponding to the base-flow velocity profile in the non-Newtonian layer (see Section III). These coefficients are used to compute the velocity derivatives and the effective and tangent viscosities needed for the stability analysis.

TABLE I. Convergence of the critical superficial velocity of the heavy phase $\left(U_{1S}, [\text{m/s}]\right)$ with increasing the truncation number $N$.

| Order of Chebyshev polynomials, $N$ | Horizontal air-CMC03 flow: $H = 0.02\,\text{m},\ r = 1000,\ m = 2662,$ $n = 0.7556,\ \lambda_C = 0.0902,\ \sigma = 0.072\,\text{N/m}$ | Horizontal oil-in-water emulsion/water flow: $H = 0.02\,\text{m},\ 1/r = 1.25,$ $m = \mu_0/\mu_2 = 723,\ \tilde{m} = \mu_\infty/\mu_2 = 18.1,$ $n = 0.4341,\ \lambda_C = 0.4583,\ \sigma = 0.03\,\text{N/m}$ |
|---|---|---|
| | $U_{2S} = 0.013\,\text{m/s},\ k = 0.1$ | $U_{2S} = 0.0528\,\text{m/s},\ k = 0.5$ |
| 40 | 0.924102 | $1.9991 \cdot 10^{-2}$ |
| 50 | 0.924103 | $1.99909 \cdot 10^{-2}$ |
| 60 | 0.924103 | $1.99909 \cdot 10^{-2}$ |
| 70 | 0.924103 | $1.99909 \cdot 10^{-2}$ |

## V. RESULTS AND DISCUSSION

The goal of this study is to obtain stability limits of smooth-stratified flow (i.e., the region that is stable with respect to all wavelength perturbations) and to reveal the destabilizing mechanisms involved. Along the stability boundary of this flow configuration all perturbations are damped, except the one with a particular wavenumber that is neutrally stable and is responsible for triggering instability (i.e., the critical perturbation).

For systems with a shear-thinning liquid as one of the phases, in addition to the seven dimensionless parameters governing the Newtonian stability problem $\left(\text{i.e., } m, q, r, \text{Re}_{2S} \text{ or Fr}_{2S}, \text{We}_{2S}, Y, \text{and } \beta\right)$, there are three parameters connected to the Carreau rheological model $\left(n, \tilde{\lambda}_C, m_\infty\right)$. $\text{Re}_{jS} = \rho_j U_{jS} H / \mu_j$ and $\text{Fr}_{jS} = U_{jS}^2 / (gH)$ are the superficial Reynolds and Froude numbers respectively, and $\text{We}_{jS} = \rho_j U_{jS}^2 H / \sigma$ is the superficial Weber number. The parameter $m_\infty$ is considered only for liquid-liquid flows, where high shear rates are achievable in the non-Newtonian layer, while it can be ignored for the studied gas-liquid systems, where such high shear rates in the liquid are not reached for flow conditions of practical interests (see Picchi et al., 2017).

Due to the large number of dimensionless parameters and for the sake of physical interpretation of the results, we prefer to deal with real shear-thinning liquids. The selected liquids are those that have been commonly used in experimental investigations and their full rheological curves are available in the literature (e.g., Sousa et al., 2005 and Partal et al., 1997). These liquids are highly viscous shear-thinning liquids and their rheological data can be found in Table II. Note that for real liquids the rheological parameters of the Carreau model are not independent: an increase of $\mu_0$ and $\lambda_C$ is typically accompanied by a decrease of $n$.

TABLE II. Carreau viscosity model parameters for different CMC solutions by Sousa *et al.* (2005)[*] and a stabilized oil-in-water emulsion by Partal *et al.* (1997).

| Solution | $\mu_0$ (Pa s) | $\mu_\infty$ (Pa s) | $\lambda_C$ (s) | $n$ (-) |
|---|---|---|---|---|
| Water-CMC03 | 0.0484 | 0.0 | 0.0902 | 0.7556 |
| Water-CMC08 | 0.9831 | 0.0 | 0.4639 | 0.5116 |
| 3%SE 60% O/W Emulsion (35°) | 0.7230 | 0.0181 | 0.4583 | 0.4341 |

[*] The data by Sousa et al. (2005) is here fitted using the Carreau model instead of the Carreau-Yasuda model.

Owing to complexity of the current problem, the following working methodology is suggested. At first, due to dependence of the effective viscosity on the superficial velocities, the base (steady state) flow "rheological regions" (Picchi et al., 2017) are depicted on the flow pattern map. In the "Newtonian region", the viscous effects of the non-Newtonian phase are dominant, and the base-flow integral variables (holdup and dimensionless pressure gradient) can be predicted by assuming a Newtonian liquid with a viscosity corresponding to the zero-shear-rate viscosity of the shear-thinning liquid. In the "shear-thinning region", the shear-thinning behaviour of the non-Newtonian phase significantly affects the two-phase flow characteristics. Thus, it may happen that smooth stratified flow is stable only within the Newtonian region, in which case the stability analysis may be reduced to the simpler Newtonian stability problem.

For this reason, as the second stage, the analytical solution for the long-wave stability boundary for the two Newtonian fluids (corresponding to the viscosity ratio *m*) is depicted on the flow pattern map. If this boundary is within the Newtonian region, it provides an upper bound for the stable region. Depending on whether the obtained long-wave stability boundary is in the Newtonian region or not, the stability analysis with respect to all wavelength perturbations is carried out for the Newtonian or non-Newtonian problem, respectively.

In the following, we apply this procedure to several representative cases. The exact results for all wavelength stability analysis, obtained by considering the Carreau viscosity model for the non-Newtonian liquid, are presented on the flow pattern map. The stability limits for stratified flow corresponding Newtonian fluids (with the same $m$) are also demonstrated for the sake of interpretation and validation of the above-suggested procedure. Moreover, the critical perturbation along the stability boundary (i.e., which is responsible for triggering instability and described by the leading eigenfunction) is reported and its pattern is discussed. These results allow us to make some additional conclusions on the nature of the instability. Along with the critical perturbations, the profiles of effective $(\mu)$ and tangent $(\mu_t)$ viscosities are provided to show the effect of the rheological behavior on the flow instability.

**A. Gas-Liquid horizontal flows**

Gas-liquid systems are characterized by high density and viscosity ratios, and the majority of works on non-Newtonian fluids have been devoted to gas-liquid flows. The shear-thinning liquid is the heavy phase, which is usually a solution of a polymer (e.g., Carboxymethyl cellulose or Xanthan gum) and water, while air is considered to form the light Newtonian phase. The air density and viscosity are $\rho_2 = 1 [\text{kg/m}^3]$ and $\mu_2 = 1.81 \cdot 10^{-5} [\text{Pa} \cdot \text{s}]$, respectively.

Considering the air-CMC03 flow as an example, the heavy phase (CMC03) is much more viscous $(m = 2662)$ than the light phase (air), and the density ratio $(r = 1000)$ is much higher than in liquid-liquid systems. The rheological properties of CMC03 are shown in Table II. Note that when referring to the Newtonian stability boundary, a Newtonian liquid of viscosity $\mu_0$ is considered. The stability map for horizontal flow in the channel of 2 cm height is shown in Fig. 2. The stability boundaries are plotted in the superficial velocities coordinates, where x-coordinate is the light (air) superficial velocity $U_{2S}$ and y-coordinate is the heavy (CMC03) superficial velocity $U_{1S}$. The value of the surface tension is taken as $\sigma = 0.072 [\text{N/m}]$ (typical value for water-CMC solutions, see Picchi et al., 2015). As described in Picchi et al. (2017), this system has a relatively large Newtonian region, depicted by the black dash-dotted line in Fig. 2. In this region the predicted liquid (heavy phase) holdup computed by using Carreau model is practically the same as that obtained for Newtonian liquid with the zero-shear rate viscosity $(\mu_0)$. The (Newtonian) long-wave neutral stability boundary is located within the Newtonian region except for high holdups (e.g., $h > 0.95$, the iso-holdup line is depicted in Fig. 2 as a dashed line). Note that for high air and liquid superficial velocities, the long-wave stability boundary approaches the zero-gravity stability boundary, i.e., a line of the critical flow rate ratio of $q_{cr} = (U_{1s}/U_{2s})_{cr} = \sqrt{m} = 51.6$, where the interfacial shear is zero and the average velocities of the two phases are equal (see Kushnir et al., 2014). The stable region for long waves is unbounded for low air superficial velocities.

Taking into account all wavelength perturbations for operating conditions with holdup less than 0.95, the long waves are observed to be the critical ones all along the stability boundary except for small region of intermediate holdups (the critical wavenumbers normalized by the channel height, $k_H = 2\pi H / l_{wave}$, are shown in Fig. 2). The results indicate that, in fact, the stability boundaries for stratified flows of Newtonian fluids and Newtonian/non-Newtonian (Carreau) shear-thinning fluids coincide within the Newtonian region. At high liquid holdups, the shear-thinning effect is important, and as expected, the all wavelength stability boundary for Carreau liquid differs from the one for Newtonian fluids. With such a viscous liquid, maintaining stratified flow for very high holdup (e.g., $h > 0.95$) would be difficult, since even small perturbations at the liquid entrance or downstream would result in bridging of the channel and blockage of the air flow leading to flow pattern transition. Therefore, from the practical point of view, the details of the predicted stability boundary in the $h > 0.95$ region are of limited interest. Nevertheless, for the sake of completeness, the results obtained at the high holdup region are presented for this case study. As shown, the resulting stability boundary consists of two branches: the left branch corresponds to short-wavelength perturbations, while along the right branch long-wavelength perturbations were found to be responsible for triggering instability. This long-wave boundary approaches the zero-gravity stability boundary of the Carreau liquid, which however cannot be obtained analytically. Due to the shear-thinning effect, it is obviously located below the one corresponding to the $\mu_0$-Newtonian (zero-viscosity) limit (i.e., below $q_{cr} = \sqrt{m}$). In the narrow stable strip confined by these two branches, the long waves are stable, and the short-wave perturbations are stabilized by surface tension. In fact, gravity effects are negligible in this region, since it corresponds to very high superficial velocity of the viscous liquid (and the air). Such a stable narrow region was predicted also for a zero-gravity air-water system (see Barmak et al., 2016a). Note that the operational window of the experiments conducted by Picchi et al. (2015) is beyond the obtained stability boundary, and indeed, stratified flow was not observed.

Examining the patterns of the critical perturbations (eigenfunctions) for three characteristic points (**A**, **B**, and **C** marked in Fig. 2) gives an additional physical insight into the flow destabilization mechanisms. The absolute values of the eigenfunction amplitude $\left(|\phi|\right)$ are shown in Figures 3(a), (c), (e) (red solid lines). For the purpose of further discussion, the base-flow velocity profiles (green dashed lines) are also depicted in these figures, since in single-phase flow the location of the maximal $|\phi|$ coincide with that of the maximal base-flow velocity. All the profiles are normalized by the corresponding value at the interface. Note that $|\phi'|$ is discontinuous across the interface and is normalized by $|\phi'_2(0)|$ in the Newtonian phase. The effective $(\mu)$ and tangent $(\mu_t)$ viscosities as a function of the cross-section coordinate $y$ in the non-Newtonian layer $(-h_2 \leq y \leq 0)$ are provided in Figures 3(b), (d), (f) to show the degree of shear-thinning effect.

Assuming that the maximum of the perturbation amplitude corresponds to the location where the instability evolves in the flow, it is possible to make some speculations on the flow destabilization. Depending on the flow conditions, the critical perturbation can evolve mainly at the interface (so-called "interfacial mode" instability) or in the bulk of one of the phases (i.e., "shear mode"). As seen, for high liquid holdups (points **A** and **B**), the instability can be associated with an interfacial mode, although the critical perturbation is a short wave at point **A** and a long wave at point **B**. For point **A**, the base-flow velocity profile also exhibits a maximum at the interface. However, at point **B**, for a lower liquid superficial velocity (but almost the same holdup) the maximum of the base-flow velocity is in the bulk of the air layer. At lower liquid holdup (point **C**) the maximal perturbation is in the bulk of the air layer implying that the critical disturbance is associated with a shear mode of instability. The maximal $|\phi|$ is shifted towards the interface as compared to the location of the maximal base-flow velocity. For even lower holdups (i.e., lower liquid superficial velocities), the critical perturbation corresponds to a long wave, and its maximal $|\phi|$ approaches the location of the maximum of the base-flow velocity. In fact, under such conditions the highly viscous liquid layer is practically like a wall for the fast air layer (even though the liquid still occupies a bit less than half a channel).

The switch between two modes of instability occurs around the holdup $h = 0.81$ $\left(U_{1S} = 0.025\,\text{m/s}; U_{2S} = 0.02\,\text{m/s}\right)$, where a maximum of the perturbation amplitude in the air phase becomes dominant.

As expected for the shear-thinning fluid, the values of the tangent viscosity are always lower than those of the effective viscosity $(\mu_t < \mu)$. Since at points **A**-**C** the shear rate in the non-Newtonian liquid at the interface is very low, $\mu_t$ and $\mu$ converge to 1 (i.e., zero-shear-rate viscosity Newtonian behavior). This is typical for gas-liquid flows, where the gas shears the heavier viscous liquid layer. The shear-thinning effect is more pronounced at point **A** (i.e., the variation of $\mu_t$ and $\mu$ is more significant), as this point is located deeper in the shear-thinning region compared to point **B**. As expected, at all three points, the viscosity attains minimal values at the wall where the shear rate is maximal. However, within the Newtonian region (point **C**), the (dimensionless) viscosity of the shear-thinning liquid only slightly deviates from 1.

The stability map for the horizontal air-CMC08 flow is presented in Fig. 4. Compared to the air-CMC03 considered above, this system has a higher viscosity ratio $(m = 54071)$ and the shear-thinning liquid behavior index is lower $(n = 0.5116)$. Although the Newtonian region diminishes in this case, the increase of the liquid viscosity results in a shift of the long-wave stability boundary to lower liquid superficial velocities, and the stability boundary is almost completely (except for high holdups) within that region. Long waves are the critical perturbations for the entire range of holdups. However, a small discrepancy between the Newtonian and non-Newtonian stability boundaries can be noticed, due to the proximity of the stability boundary to the shear-thinning region.

Instability of thick liquid layers (e.g., $h > 0.5$) may result in transition to elongated-bubble or slug flow, in particular when the instability is associated with an interfacial mode (e.g., point **A**, **B** in Fig. 3). For thin liquid layers, the instability is expected to be associated with transition to stratified wavy flow. In the above test cases, with a channel size of $H = 0.02\,\text{m}$, long waves trigger the instability of thin liquid layers. For this channel size, the maximal air superficial velocity for maintaining smooth stratified flow (denoted as the critical air superficial velocity) is found to be $U_{2S} \approx 6.1\,[\text{m/s}]$. For air-CMC03 flow, the stability boundary almost attains this velocity (see bottom right corner of Fig. 2), while for the more viscous air-CMC08 system the critical air superficial velocity corresponds to lower water superficial velocities (not shown in Fig. 4). For systems of even higher liquid viscosity (higher $m$), the same maximal critical gas superficial velocity would be obtained, however, at even lower liquid superficial velocity (i.e., lower $q$).

The effect of the channel size on the stability boundary is of obvious interest, in particular for upscaling lab-scale (and low pressure) data on the stratified-smooth boundary to field operational conditions, typically involving large channels and high pressure (and consequently higher gas density and lower $r$). The upscaling and downscaling rules for predicting the effect of channel size and gas pressure on the maximal critical gas flow rate were formulated and discussed in Barmak et al. (2016a) with respect to gas-water (i.e., low viscosity liquid) systems. The obtained scaling rules have been revalidated here for their applicability to the gas/shear-thinning very viscous liquids that are considered in this study. In fact, those scaling rules are found to be valid as the maximal gas velocity is associated with a Newtonian behavior of the liquid layer.

Upon reducing the channel size from 0.02m, the low holdup part of the stability boundary is associated with long-wave instability. In this case (see Kushnir et al., 2014, 2017) the critical air superficial velocity is found to correspond to a critical gas superficial Froude number $(\text{Fr}_{2S}^{Cr} = U_{2S} \cdot \left[(r-1)gH\right]^{-0.5} \approx 0.45)$, whereby this critical velocity decreases with reducing the channel size $(U_{2S} \propto H^{0.5})$. Note that for highly viscous liquids the critical gas Froude number is somewhat higher than that of the air-water flow ($\approx 0.35$, see Barmak et al., 2016a). For example, if the channel size is reduced from 0.02m to 0.002m, the maximal critical gas superficial velocity decreases to $1.9\,[\mathrm{m/s}]$. Although the range of air superficial velocities where the liquid exhibits a Newtonian behavior diminishes as well with reducing the channel size (see Picchi et al, 2017), the critical air velocity is still well within the Newtonian region.

With increasing the channel size, the Newtonian region expands, but the critical perturbations may no longer correspond to long waves. In fact, when the channel size is increased to 0.2m, the critical disturbances for thin liquid layers correspond to short waves. In this case the critical gas superficial velocity corresponds to a critical superficial Reynolds number $\left(\mathrm{Re}_{2S}^{Cr} = \rho_2 U_{2S} H / \mu_2 \approx 7800\right)$, whereby increasing the channel size results in a decrease of the critical air superficial velocity $(U_{2S} \propto 1/H)$. Hence, the stability boundary of low liquid holdups is shifted deeper into the Newtonian region. Note that the critical Reynolds number is the same as that obtained in Barmak et al. (2016a) for air-water system and is practically insensitive to the gas-liquid viscosity ratio. This is obviously due to the fact that for thin layers of very viscous liquid, the interface practically represents a solid surface for the gas flow.

The widest stable region for air-CMC solutions flow is obtained in $H = H_{Cr} = 0.022\,\mathrm{m}$ channel, where the critical air velocity corresponds both to the above values of $\mathrm{Fr}_{2S}^{Cr}$ and $\mathrm{Re}_{2S}^{Cr}$, and the long waves $(k \to 0)$ are still the critical perturbation for thin water layers. In this case the critical air superficial velocity for air-CMC stratified flow is maximal, $U_{2S}^{Cr} = 6.4\,\mathrm{m/s}$ (corresponding to $h \approx 0.1$), whereby the stratified-smooth region extends over the largest range of superficial air velocities. The above results are in agreement with the formulas for the critical channel height and the corresponding maximum of the critical gas superficial velocity presented in Barmak et al. (2016a): $H_{Cr} \approx \left[\left(\mathrm{Re}_{2S}^{Cr}\right)^2 \mu_2^2 \Big/ \left(\left(\mathrm{Fr}_{2S}^{Cr}\right)^2 \rho_2^2 g(r-1)\right)\right]^{1/3}$ and $U_{2S}^{Cr} = \left[(r-1)gH_{Cr}\right]^{1/2} \mathrm{Fr}_{2S}^{Cr}$. With increasing the gas pressure (i.e., $\rho_2 > 1\,[\mathrm{kg/m^3}]$ and $r < 1000$) the maximal critical gas superficial velocity decreases $(U_{2S}^{Cr} \propto p^{-2/3})$, and is obtained in smaller channel $(H_{Cr} \propto p^{-1/3})$. It is worth noting, that the effect of changing the value of $r$ by varying the liquid density ($\rho_1$) on $U_{2S}^{Cr}$ is not equivalent to that resulting from a corresponding change in the gas density $(U_{2S}^{Cr} \propto \rho_1^{1/3}$ versus $U_{2S}^{Cr} \propto \rho_2^{-2/3}$ for the same $r \gg 1)$. On the other hand, for low gas superficial velocities, the critical liquid superficial velocity, $U_{1S}^{Cr}$, is practically not affected by the gas density (which is much lower than the liquid density). In fact, under such conditions the liquid flow behaves like a single-phase liquid flow in an open channel (the main effect of the thin gas layer is the detachment of the interface from the channel wall).

The general conclusion to be drawn from the case studies considered here is that the stability of horizontal stratified flow of gas/highly viscous shear-thinning liquid (in channels of $H \geq 2\,\mathrm{mm}$) can be satisfactory predicted by Newtonian liquid with zero-shear rate viscosity instead of solving the complete non-Newtonian problem. This also allows applying the same downscaling and upscaling rules for the effect of the channel size and system pressure on the critical air superficial velocity that are found valid for Newtonian gas-liquid horizontal flows.

**B. Gas-Liquid inclined flows**

The stability map obtained for the air-CMC08 flow in $\beta = 5°$ downward inclined channel is shown in Fig. 5. In downward inclined flow, the gravitational force accelerates the liquid layer, enhances the shear-thinning effect and facilitates the flow of such a highly viscous liquid (see Picchi et al., 2017). Compared to horizontal flows, the Newtonian region shrinks to very low liquid superficial velocities, and the entire range of practical superficial velocities belongs to the shear-thinning region. The dominancy of gravity is manifested by the almost horizontal iso-holdup lines in a wide range of operational conditions, except at high liquid and gas superficial velocities where gravity effects are negligible and the iso-holdup lines approach those obtained in horizontal flow.

Assuming Newtonian liquid of viscosity $\mu_0$, the stability analysis indicates that long waves are the critical perturbations and the long-wave analytical stability boundary (violet dashed line) coincides with the all wavelength boundary (blue solid line). Surprisingly, for such a viscous liquid the channel inclination results in stabilization of the flow, whereby the stable region extends to higher liquid superficial velocities compared to horizontal flow (Fig. 4). This is in contrast to the results obtained for less viscous air-water system, where an increase of inclination has a destabilizing effect (Barmak et al., 2016b). In fact, for such conditions the flow destabilization is dominated by the liquid inertia. Accordingly, the critical liquid superficial velocity corresponds to a critical liquid Froude number $\left(\text{Fr}_1^{Cr} = \left[U_{1S}/h \cdot \left(g\cos(\beta)Hh\right)^{-0.5}\right]_{Cr}\right)$, which is found to be independent of the liquid viscosity and corresponds $\text{Fr}_1^{Cr} \approx 0.53$ (for $\beta = 5°$ and $H = 0.02\,\text{m}$). Consequently, the higher is the liquid viscosity, the higher is its critical superficial velocity (and the corresponding holdup) for the flow destabilization. However, for a very high liquid viscosity (e.g., $\mu_0$ of the CMC08 solution) the critical holdup (corresponding to the critical $U_{1s}$) is extremely high $(h > 0.95)$, where the stratified flow would be practically infeasible, and transition to slug flow may take place. Since as the liquid viscosity increases, the (practical) threshold on the holdup $(\approx 0.9)$ for the stratified-wavy/slug transition is reached at lower $U_{1s}$, the stratified-wavy region is expected to shrink with the increase of the liquid viscosity. However, for down flow of the air/CMC-solution systems, the Newtonian stability limits are deep inside the shear-thinning region. Therefore, the stability problem for a Carreau liquid should be considered to obtain correct results for these two-phase systems.

The results of the stability analysis show (Fig. 5) that, indeed, the non-Newtonian stability boundary differs from the Newtonian one owing to the shear-thinning behavior, which affects both the base flow and the perturbed flow characteristics. Still, long waves are found to be the critical perturbation all along the stability boundary. The non-Newtonian and Newtonian stability boundaries converge only for low holdups and high air superficial velocities. The critical air superficial velocity for thin liquid layer is identical $(U_{2S} \approx 6.1\,[\text{m/s}])$ to that obtained in horizontal flow and corresponds to the same critical Froude number ($\text{Fr}_{2S}^{Cr} \approx 0.45$). At low air superficial velocities the stability boundary corresponds to a constant liquid superficial velocity $(U_{1S} = 0.064\,[\text{m/s}])$ and holdup of $h = 0.56$ (corresponding to $\text{Fr}_1^{Cr} \approx 0.35$), which are lower than that obtained by assuming a Newtonian liquid. Yet, the above discussion on the effects of the $\mu_0$ viscosity on the critical non-Newtonian liquid velocities (for stratified smooth/wavy and stratified/slug transitions) are still valid. These findings are in accordance with the experimental results of Picchi et al. (2015), where only stratified-wavy flow pattern was observed as the tests were conducted for higher liquid superficial velocities than those predicted by the stability analysis. Moreover, the stratified wavy region was found to shrink with the increase of the liquid viscosity.

Due to competition between gravity and shear forces multiple base-flow configurations (i.e., up to three solutions for the holdup) can be encountered in downward inclined gas/shear-thinning liquid flows (Picchi et al., 2016a, 2017). The triple-solution (3-s) region in downward air-CMC08 flow is obtained at high liquid superficial velocities and relatively low air superficial velocities. The stability analysis (for a Carreau liquid) indicates that within the 3-s region, the lower holdup solution becomes stable in a narrow range of high liquid superficial velocities, while the middle and upper solution are always unstable. Despite the fact that the flow is predicted to be stable also at higher gas superficial velocities in the proximity of the 3-s region, for such high holdups of the viscous liquid in this region stable stratified flow is anticipated to be practically unfeasible.

Figure 6 shows the patterns of the critical (long-wave) perturbations along with the base-flow velocity profiles (Figures 6(a), (c), (e)) at several representative points (**A**-**C** in Fig. 5) along the stability boundary. Points **A** and **B** correspond to the same liquid superficial velocities and holdup. At point **A**, the air is dragged downward by the viscous liquid, and the maximal velocity is at the interface, while backflow (up-flow) of the air phase occupies the zone adjacent to the upper wall. The air layer flows faster than the liquid layer at point **B**, where the maximum of the base-flow velocity profile is in the bulk of the air. The perturbation patterns are also different: at point **A**, the eigenfunction maximum is within in the air layer, implying that the critical perturbation can be associated with a shear mode of instability, while at point (**B**) the critical perturbation is clearly associated with an interfacial instability. At even higher air superficial velocity and lower holdup (point **C**), the air flows much faster than the liquid and the critical instability clearly corresponds to a shear mode. The shear-thinning behavior of the liquid phase at all the tested points becomes evident by examining the (dimensionless) viscosity profiles (see Figures 6(b), (d), (f)). Nevertheless, the existence of the $\mu_0$ viscosity limit cannot be ignored as it dominates the interaction of the two-phases at the interface and, consequently, the characteristics of the perturbed flow.

Additional insight on flow instability can be obtained by examining the real part of the stream function of the critical perturbation (Eq. (16)) at a particular time (e.g., $t=0$). It is defined as

$$\text{Re}(\psi_j) = \text{Re}(\phi_j(y)e^{ikx}) = \text{Re}(\phi_j(y)) \cdot \cos(kx) - \text{Im}(\phi_j(y)) \cdot \sin(kx). \tag{25}$$

Figure 7 shows the 2D contours of the critical perturbations ($\text{Re}(\psi)$, Figures 7(a), (c), (e)) and of the critically perturbed flow ($\text{Re}(\psi_{flow})$, Figures 7(b), (d), (f)) for conditions corresponding to points **A**-**C** in Fig. 6. Note that although infinitely long waves $(k \to 0, l_{wave} \to \infty)$ are the critical perturbations in the flow, in practical applications the channel is of a finite length. Hence, for demonstrating the perturbed flow, a perturbation of a shorter wavelength is selected ($k = 0.01$, at near critical conditions), whose growth is still close to zero, and the shape of streamlines and disturbed interface are similar to those of longer wave perturbations (see details in Barmak et al., 2016b). It is important to mention that the flow is not steady with respect to the stationary frame of reference (which is used in the present study), and the (moving) interface at a particular moment of time may not coincide with a streamline (Figures 7(b), (d), (f), see details in Barmak et al., 2016b).

As shown in Figures 7(a), (c), (e), the perturbations are represented by pairs of antisymmetric vortices (of $l_{wave}/2$ width) with their core, i.e., the maximum of the stream function perturbations, located in the bulk of air layer (**A** and **C**) or at the interface (**B**). Due to the backflow in the air layer at point **A**, a generation of the circulation cells is observed in the perturbed flow (Fig. 7(b)), whose centers lie on the line of zero base state velocity. Although the location of the maximal stream function amplitude is within the air layer, the deformation of interface is still significant. At point **B**, the largest amplitude is observed at the vicinity of the interface (Figures 7(c) and (d)). At point **C**, where the liquid base-flow velocity is very small compared to the air velocity, even small perturbations may lead to formation of vortices near the wave crests (Fig. 7(f)), which may lead to liquid entrainment into the air.

We tested also the stability of gas/shear-thinning liquid stratified flow in upward inclined channels. Although the results are not shown here, it is worth noting that in this case the long wave stability boundary corresponds to very high holdups and is found to be deep within the Newtonian region. Therefore, the complete stability analysis reduces to the simpler Newtonian problem (Barmak et al., 2016b). For upward inclined gas-liquid flows, relatively low holdup solutions are obtained only in the triple solution region (and its vicinity). However, for the highly viscous liquids considered in this study, the triple-solution region is associated with extremely low liquid superficial velocities (see Picchi et al., 2017). The conclusion is that smooth stratified flow in upward inclined channels is not feasible in the practical range of flow rates, which is associated with very high holdups. For such conditions, the instability would result in slug flow as was experimentally observed (Picchi et al., 2015).

### C. Two-fluid model for gas-liquid flows

Two-Fluid models are widely used to study the stability of gas-liquid two-phase flows. In this approach, the stability analysis is carried out based on the one-dimensional transient Two-Fluid model equations, where long-wave perturbation is an inherent assumption in the model. It was shown (subsection A) that even though the heavy phase is a very viscous shear-thinning liquid, in horizontal gas-liquid flows the stable region of smooth stratified flow is confined to the Newtonian region for the practical range of interest. This allows us to examine the applicability of stability criteria obtained for Newtonian fluids via the Two-Fluid model as a prediction tool for these operational conditions.

Since closure relations for the wall and interfacial shear stresses are required in the Two-fluid model, the common approach is to assume that these are adequately represented by quasi-steady models (based on the local holdup). Consequently, in the stability analysis, only the components of the stresses in phase with the wave height are considered. However, it was shown that the wave-induced shear stresses, which are in phase with the wave slope, should not be ignored even in the framework of long-wave assumption (for details see, e.g., Brauner and Moalem Maron, 1993, Kushnir et al., 2014). When these modifications are included in the Two-Fluid model, the resulting neutral stability criterion reads (for details see, e.g., Kushnir et al., 2007, 2017):

$$J_1 + J_2 + J_h = 1 \qquad (26)$$

with

$$J_1 = \left(\frac{\rho_1}{\rho_1 - \rho_2}\right)\frac{U_{1S}^2}{Hg\cos\beta}\frac{1}{h^3}\left[\left(\frac{\hat{c}_R}{\bar{U}_1} - 1\right)^2 + (\gamma_1 - 1)\left(1 - 2\frac{\hat{c}_R}{\bar{U}_1}\right) + \Delta\gamma_1\right], \qquad (27)$$

$$J_2 = \left(\frac{\rho_2}{\rho_1 - \rho_2}\right)\frac{U_{2S}^2}{Hg\cos\beta}\frac{1}{(1-h)^3}\left[\left(\frac{\hat{c}_R}{\bar{U}_2} - 1\right)^2 + (\gamma_2 - 1)\left(1 - 2\frac{\hat{c}_R}{\bar{U}_2}\right) + \Delta\gamma_2\right]. \qquad (28)$$

The $J_1$ and $J_2$ terms represents the (destabilizing) inertia of each phases relative to the (stabilizing) gravity (i.e., the Kelvin-Helmholtz (K-H) mechanism), while $J_h$ is the so-called 'sheltering' term, which is responsible for the destabilizing effect due to wave-induced tangential (wall and interfacial) shear stresses in phase with the wave slope. The exact values of the inertia terms $(J_{1,2})$ can be obtained based on the exact solution of the base flow. Note that the holdup ($h$), the average phase velocities $(\bar{U}_1, \bar{U}_2)$ and the long-wave velocity $(\hat{c}_R)$ can be exactly reproduced by using the MTF model closure relations for the base-flow shear stresses (Ullmann et al., 2004), while the (exact) values of the velocity profile shape factors $\gamma_1, \gamma_2$ and terms evolving from their derivatives $\Delta\gamma_1, \Delta\gamma_2$ require the base-flow velocity profiles. While the exact value of the $J_h$ term can be determined only based on the exact long-wave stability analysis, the following simple closure relation for this term has been recently formulated by Kushnir et al. (2017):

$$J_h = C_h \frac{\text{Fr}_{2S}^2}{h(1-h)^3(r-1)\cos(\beta)}\left(1 - \frac{q(1-h)}{h}\right), \qquad (29)$$

where $C_h$ is the apparent sheltering coefficient. A method to obtain $C_h$ based on the exact long-wave solution was provided by Kushnir et al. (2017), which is shown to be generally dependent on the viscosity ratio, the density ratio and the holdup, $C_h = C_h(m, r, h)$. However, for $m \gg 1$ the coefficient $C_h$ reaches a constant asymptotic value $(C_h = 0.257)$, which was found to be only mildly dependent on $h$ and $r$. It is of interest to examine the results obtained by applying the above TF stability criterion (Eq. (26)) with this constant $C_h$ value for the high viscosity ratio Newtonian systems considered in the present study. This is also relevant for horizontal gas/shear-thinning liquid flows (in the Newtonian regions).

The comparison between the stability boundaries for horizontal air-CMC03 and air-CMC08 flows predicted by the Two-Fluid model and the exact all wavelength stability boundary (present work) is depicted in Figures 8 and 9, respectively. As shown, for low liquid holdups, where instability is expected to result in a transition to stratified-wavy flow, in both cases the Two-fluid model (with the constant $C_h$ value) is able to predict correctly the critical superficial air velocity corresponding to this transition (see the bottom right corner in Figures 8 and 9). The overprediction of the stable region for $0.8 < U_{2S} < 5 [\text{m/s}]$ is because in this range the critical perturbations are short waves. Considering the air-CMC08 case, where the exact stability boundary coincides with long-wave boundary in the whole range of operational conditions of interest, the Two-Fluid model boundary follows the exact boundary up to the region of high holdups. The Two-Fluid stability boundary obtained by considering only the inertia terms (i.e., the K-H mechanism) is also presented, showing that by ignoring the destabilizing effect of the sheltering term, $J_h$, the stable smooth-stratified flow region is overpredicted.

In order to shed light on the dominant destabilization mechanism, the values of the $J$ terms as a function of the holdup along the Two-Fluid stability boundary are presented in Fig. 10. The sum of the $J$ terms is equal to 1 on the neutral stability boundary (see Eq. (26)), and their relative contributions indicate whether the instability is caused by the inertia terms $J_1$ or/and $J_2$ (KH mechanism), or by $J_h$ (sheltering mechanism), or by combination of these two mechanisms. As shown in Fig.10, in stratified air-CMC08 flow, the sheltering mechanism is dominant at low holdups, while a combination of sheltering and the air inertia is responsible for the instability up to $h \approx 0.8$. Only at higher holdups, the stability is controlled by the inertia of the viscous liquid (i.e., liquid-dominated KH instability). The difference between the sheltering terms calculated based on an asymptotic value of $C_h$ (=0.257) and its exact value that evolves form the analytical (exact) long wave stability analysis ($J_h$ and $J_{h\_exact}$ in Fig. 10, respectively) is small for most of the holdup range, and it becomes significant only at very high holdups, that leads to higher discrepancy in the stability results.

Thus, although a relatively simple model, the Two-Fluid model for Newtonian fluids (i.e. considering the liquid as a Newtonian phase with zero-shear-rate viscosity) can be used to predict the stability of horizontal stratified smooth flow. Moreover, this model gives us a tool to quantify the relative importance of the different mechanisms involved in the flow destabilization.

### D. Liquid-Liquid horizontal flow

A typical liquid-liquid system is the flow of a very viscous shear-thinning emulsion (or waxy oil) lubricated by water. Emulsions usually exhibit a complex rheology (e.g., see Partal et al., 1997), and the rheological parameters of oil-in-water emulsion are presented in Table II. It is important to emphasize that for this liquid the infinity shear rate viscosity $\mu_\infty$ cannot be neglected, since emulsions behave like a Newtonian (constant-viscosity) liquid at high shear rates. In the considered stratified flow the non-Newtonian oil-in-water emulsion is the light phase (denoted as "1") and the water is the heavy phase (denoted as "2"), which actually forms the lower layer. Therefore, as mentioned in the problem formulation (Section II), the gravitational force in this case points from phase "1" to phase "2" (i. e., $g < 0$, the opposite direction from gas-liquid flows shown in Fig. 1).

For better interpretation of the results, we first refer to the stability of two limiting Newtonian systems: one is defined by $m = \mu_0/\mu_2 = 723$ and the other by $m_\infty = \mu_\infty/\mu_2 = 18.1$ (the corresponding viscosity ratio between the heavy water layer and the lighter emulsion is $1/m = \mu_2/\mu_0 = 1.383 \cdot 10^{-3}$ and $1/m_\infty = \mu_2/\mu_\infty = 0.055$, respectively). The stability boundaries for these two Newtonian flows in the channel of 2 cm height are presented in Fig. 11. The density ratios are identical in both cases and correspond to a shear-thinning liquid/water value, $1/r = \rho_2/\rho_1 = 1.25$.

The water density and viscosity are $\rho_2 = 1000 [\text{kg/m}^3]$ and $\mu_2 = 0.001 [\text{Pa} \cdot \text{s}]$, respectively, and the surface tension is $\sigma = 0.03 [\text{N/m}]$.

The long-wave stability boundary for the flow of water and very viscous Newtonian liquid ($\mu_0$-viscosity) consists of two separate branches (violet dashed line in Fig. 11). It practically coincides with the zero-gravity long-wave stability boundary in the whole range of interest. This reflects the prevalence of viscosity with respect to gravity for such a high viscosity ratio and density ratio of the order of 1. On the other hand, with a Newtonian liquid of the $\mu_\infty$-viscosity, the gravity effects are significant and only a single long-wave stability branch is obtained (dash-dotted line). Considering the all wavelength stability boundary, short-wave instability is found to further limit the stable region in both cases (with $\mu_0$ - below the blue solid line, with $\mu_\infty$ - below the green dashed line). Single-phase stability limits for laminar flow of water and laminar flow of a Newtonian emulsion of the $\mu_\infty$-viscosity are also plotted in Fig. 11. The single-phase stability limit for the $\mu_0$-viscosity is beyond the range of superficial velocities shown in the figure. These limits are determined by the critical Reynolds number of $\text{Re}_{Cr} = 5772$, and correspond to a critical wavenumber $k_H = 1.02$ (see, e.g., Orszag, 1971). It can be easily observed that in both cases the two-phase flow is more unstable than its single-phase counterpart is.

The all wavelength stability boundary for the Newtonian/non-Newtonian (shear-thinning) liquid-liquid flow is shown in Fig. 12 (red solid line) along with the all wavelength stability boundaries for the two Newtonian limiting cases discussed above (Fig.11). The base-flow $\mu_0$-Newtonian region and the shear-thinning region are also indicated (below and above the black dash-dot line, respectively). At low emulsion superficial velocities, the non-Newtonian stability boundary practically coincides with the Newtonian one, with long waves being the critical perturbation. However, as the shear-thinning region is approached these two boundaries diverge. The shear-thinning effect tends to somewhat extend the stable region to higher superficial water velocities compared to the $\mu_0$-Newtonian boundary (except between points **B** and **C**). At high emulsion superficial velocities, the non-Newtonian stability boundary appears to be shifted from the $\mu_0$-stability boundary towards that obtained for $\mu_\infty$ (top right corner of the map). In fact, the base-flow $\mu_\infty$-Newtonian region is reached at extremely high emulsion and water superficial velocities (outside the presented map), and the stable region does not actually extend to that region. It is worth noting that for emulsion holdups $h < 0.9$, the critical perturbations are mainly long waves, except for the small region around point **B**. The region of very high holdups $(h > 0.98)$ is not shown since the stratified flow is considered practically unfeasible for such conditions.

In view of the rather complex shape of the stability boundary, it is of interest to examine the growth rate of the waves spectrum upon increasing the emulsion superficial velocity at a constant superficial water velocity (e.g., $U_{2S} = 0.01 [\text{m/s}]$). As shown in Fig. 13 (a), long-wave perturbation is critical at point **C** (see Fig.12, at lower emulsion superficial velocity all wavenumber perturbations are damped). However, upon crossing the stability boundary by slightly increasing the emulsion superficial velocity (point **C₁**, Fig. 13 (a)), there is already a range of unstable perturbations, with the most unstable perturbation shifted towards intermediate wavenumbers $\left(k_{H\_\max} \approx 0.2\right)$. For even higher $U_{1S}$ the flow becomes neutrally stable at point **C₃** (see Fig. 12), where long waves are again the critical perturbations. The growth rates of perturbations as a function of wavenumber for point **C₃** and its neighboring point **C₂** in the unstable region (at slightly lower $U_{1S}$) are presented in Fig. 13 (b).

Some further physical insight on the instability mechanism can be gained by examining the profiles of the critical perturbations and those of the effective and tangent viscosities. As can be seen in Fig. 14 (a), the critical perturbation at point **A** (see Fig 12) has a maximal amplitude is in the bulk of the faster water layer, indicating shear mode of instability. Point **A** is located in the Newtonian region (Fig. 12), and, indeed, at this point the effective and tangent viscosities are almost constant and are equal to 1 (Fig. 14 (b)). At higher emulsion superficial velocity (point **B**) the critical perturbation is a short wave, which is seen to be associated with an interfacial mode, however a secondary maximum is still observed in the water layer (at the vicinity of the maximal water velocity). Point **B** is already in the shear-thinning region, as can be seen by the viscosities profiles (Fig. 14 (d)). At this point the highest shear rate and, consequently, the lowest effective and tangent viscosities are at the wall, while toward the interface the viscosity tends to the $\mu_0$ Newtonian limit. At point **C**, which corresponds to a lower water superficial velocity and a higher emulsion holdup, a long-wavelength perturbation is critical and is associated with an interfacial mode (Fig. 14 (e)), although the maximum of the base-flow velocity profile is still in the lower layer. The shear-thinning behavior at point **C** (Fig. 14 (f)) is also similar to that at **B** due to the resemblance of the emulsion velocity profiles. The critical perturbations at points **D** and **E,** which are situated along the right branch of the stability curve, where the emulsion holdup is very high, are shown in Figures 14 (g) and (i). The base-flow velocity profiles are similar in these two cases, with a maximal velocity in the very viscous layer near the interface. This is also reflected in the emulsion (dimensionless) viscosity profiles (Figures 14 (h) and (j)), where the maximal value of 1 is attained at the vicinity of the interface (where $\dot\gamma = 0$). However, the viscosity varies significantly through the emulsion layer, indicating a strong shear-thinning behavior. The critical perturbations are short waves and correspond to an interfacial mode in both cases, while for lower water superficial velocities (point **E**) a secondary maximum in the lower layer can be observed. It is important to mention that in the Newtonian region, where the stability boundaries for the non-Newtonian and Newtonian systems coincide, the critical perturbations are also the same. On the other hand, in case the stability boundary is located in the shear-thinning region, the critical flow conditions and wavenumbers are different for the two systems, as well as the critical perturbation profiles. Obviously, in the shear-thinning region, the base-flow characteristics (e.g., holdup, velocity profile) of the two systems are different for the same flow rates. Consequently, considering for the same wavenumber, the associated instability characteristics of the critical perturbation of the non-Newtonian system differ from those obtained by assuming a Newtonian behavior. Namely, in the latter case, the perturbation is no longer critical (it can be either damped or amplified).

The results obtained for emulsion-water system (as well as for the viscous-liquid/air systems) clearly show that there is no definite correlation between the mode of instability and the perturbation wavelength. The results obtained here reinforce our previous conclusion (Barmak et al., 2016a) that long waves do not necessarily imply an interfacial mode of instability and may correspond to shear mode of instability as well. A classification to a shear mode or an interfacial mode can be made only based on examination of the pattern of the disturbance stream function.

Due to the complexity of the stability map for the above water-emulsion test case, we would like to recommend the approach we undertook for the identification of the all wavelength stability boundary for the stratified flow of Newtonian/non-Newtonian shear-thinning fluids. As shown, the search for the stability boundary and the interpretation of the results can be facilitated when it is conducted in light of the base-flow rheological map (Picchi et al., 2017) and considering as a benchmark the stability boundaries for Newtonian liquids with the limiting $\mu_0$ and $\mu_\infty$ viscosities. These can provide some rough bounds of the stability limits of the stratified flow when a non-Newtonian liquid is involved.

## VI. SUMMARY AND CONCLUSIONS

A comprehensive linear stability analysis considering all wavelength perturbations was performed for stratified flows of Newtonian/non-Newtonian shear-thinning fluids. The Carreau model has been chosen for proper modeling of the rheology of a shear-thinning fluid, as this model represents the Newtonian behavior of such fluids at low and high shear rates. The results are presented in the form of the stability boundaries on the flow map for several practically important cases. The stability maps are accompanied by spatial profiles of the critical perturbations and of the base-flow velocity. The distributions of the effective and tangent viscosities in the non-Newtonian layer are also presented to show the influence of the complex rheological behavior of shear-thinning liquids on the mechanisms that are responsible for triggering instability.

Applications of shear-thinning liquids generally involve very viscous liquids (e.g. polymer solution, emulsion, waxy oil), and the Carreau viscosity model is considered in the literature as a good rheological model for such liquids. We point out in this work that the capability of this rheological model to fit the zero- and infinity- shear-rate viscosities is a crucial aspect in obtaining the correct base-flow characteristics and the stability boundaries. In fact, in some cases the stable stratified flow is obtained only for operational conditions for which shear-thinning liquid behaves practically as Newtonian. Moreover, it was found that even in cases where the shear-thinning behavior of the liquid is prominent, the effective viscosity at the interface and its vicinity approaches the Newtonian value, while the shear-thinning behavior is exhibited mainly in the bulk and in the near-wall region of the non-Newtonian layer. This is a characteristic of the studied systems, where the shear-thinning fluid is much more viscous than the other phase, and therefore the shear rate at the interface is rather low. Obviously, a realistic prediction of the liquid viscosity at the phases' interface is essential for exploring the stability of the flow. This behavior of shear-thinning liquid in two-phase flow system can be captured only by considering a realistic rheological model (e.g., the Carreau viscosity model) and not, for example, by the simpler and widely used power-law model. Assuming a power-law fluid would result in a rigid layer (infinite viscosity) at the interface and unphysical representation of the interaction between the phases.

Since the considered problem involves many input parameters (up to ten), we proposes a working methodology to alleviate the search for the neutral stability boundary and the associated critical disturbances that are responsible for triggering the instability. The methodology enables further interpretation of the results to reveal the relative importance of viscous and shear-thinning effects. The methodology consists of a stepwise procedure, which starts by mapping rheological regions of the base flow (see Picchi et al. 2017). Then, the analytical solution for the long-wave stability of the Newtonian fluids (with the zero-shear rate viscosity of the non-Newtonian phase) is applied for identifying the locus of the corresponding boundary on the rheological base-flow map. If the resulting stable region is within the base-flow Newtonian region, the search for the all wavelength stability boundary can be carried out solving the simpler Newtonian stability problem. Otherwise, the full stability analysis of Newtonian/non-Newtonian stratified flow should be carried out. In any case, the results obtained for the stability of Newtonian fluids are also of interest, since the stability of stratified flow involving a highly viscous liquid has not been researched in the literature.

The presented test cases of gas-liquid and liquid-liquid flows demonstrate that the effects of the liquid rheology on the flow stability are rather complicated, and it is difficult to anticipate their consequences. It is shown that in many cases the investigation of the simplified Newtonian problem is sufficient for prediction of the stability boundary of smooth stratified flow (i.e., horizontal gas-liquid flows). In such cases, the knowledge gained from the stability analysis of Newtonian fluids is applicable to the non-Newtonian gas-liquid systems. In particular, the issue of scaling of the critical gas superficial velocity for low liquid holdups has been addressed. Similarly to the air-water flow (Barmak et al., 2016a), also for gas/highly viscous liquid, we found a (critical) channel size, for which the gas superficial velocity for maintaining smooth-stratified flow is maximal. For channels smaller than the critical size, the critical gas superficial velocity is scaled by a critical (superficial) Froude number. On the other hand, in large channels, the critical gas superficial velocity is scaled by a critical (superficial) Reynolds number (i.e., decreases with the channel size). These findings can be practically implemented for upscaling lab-scale data on the boundary of the stratified-smooth flow pattern to field operational conditions. Moreover, when long waves are the critical perturbation, the simple 1D Two-Fluid (TF) models developed for Newtonian systems can be used to obtain reliable prediction of the stability boundary. In particular, the recently developed TF model (Kushnir et al, 2017), which combines the K-H and 'sheltering' destabilizing mechanisms (due to wave induced shear stresses in phase with the wave slope), is shown to provide a reliable tool for the prediction of the long-waves stability boundary.

Compared to gas-liquid systems, the identification of the stability boundary of liquid-liquid systems where one of the phases is a highly viscous shear-thinning liquid is a rather tedious process. This is due to the delicate balance between the small gravity forces, surface tension and shear-thinning effects. As a test case, the stability of stratified flow of a viscous shear-thinning water-in-oil emulsion and water has been studied together with the two limiting Newtonian systems, which are characterized by the two limiting viscosity ratios $m = \mu_0 / \mu_2$ and $m_\infty = \mu_\infty / \mu_2$. These viscosity ratios are associated with low and high shear rates of the non-Newtonian liquid, respectively. Within the base-flow Newtonian rheological region (i.e., at low emulsion superficial velocities), the exact stability boundary coincides with the Newtonian one. The shear-thinning behavior has a stabilizing effect on the flow, whereby the stable region extends to higher water flow rates compared to the Newtonian $m$-system. In the shear-thinning region the exact (non-Newtonian) stability boundary is located between those obtained for the $m$- and $m_\infty$-Newtonian stability boundaries.

The considered case studies can be used as benchmark cases for future studies. In addition, the working methodology proposed in this paper facilitates the implementation of a systematic analysis for a particular system of interest.

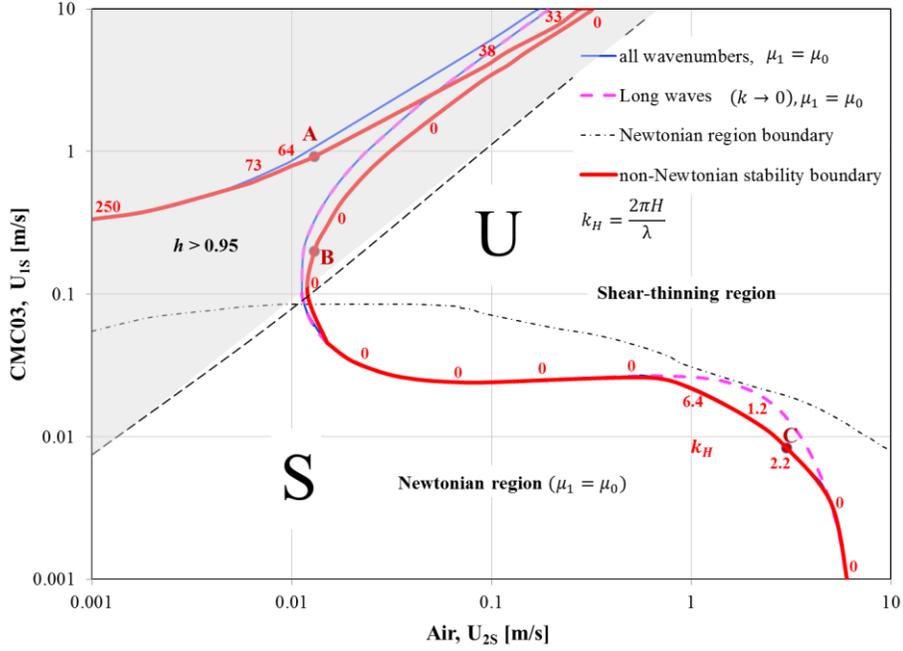

FIG. 2. Stability map for horizontal air-CMC03 flow $(r=1000; m=2662; \rho_2=1\text{kg/m}^3;$ $\mu_2=1.81\cdot10^{-5}\,\text{Pa}\cdot\text{s}; \sigma=0.072\,\text{N/m})$ in a $H=0.02\,\text{m}$ channel. $S$ and $U$ denote the stable and unstable regions, respectively. The $h=0.95$ iso-holdup line is depicted as a dashed line to show the very high holdup region.

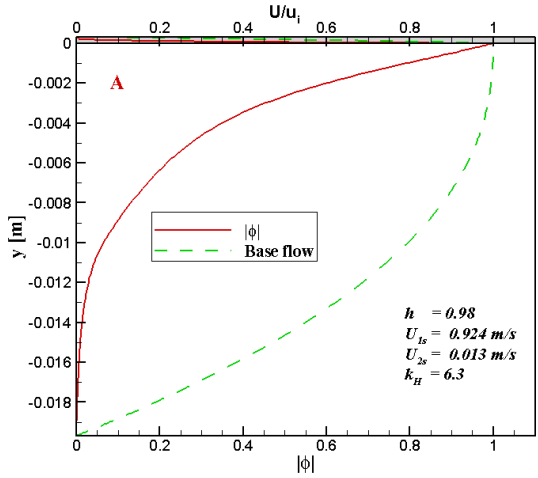

(a)

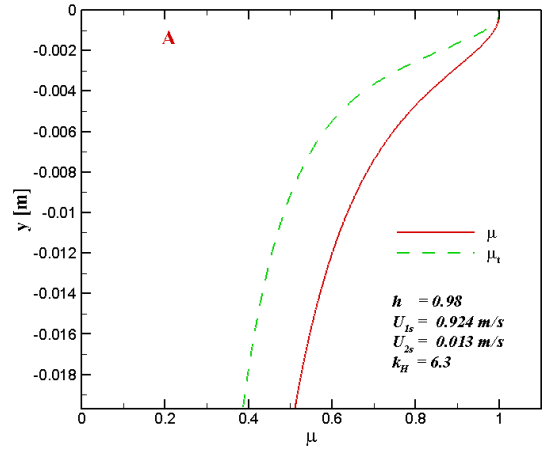

(b)

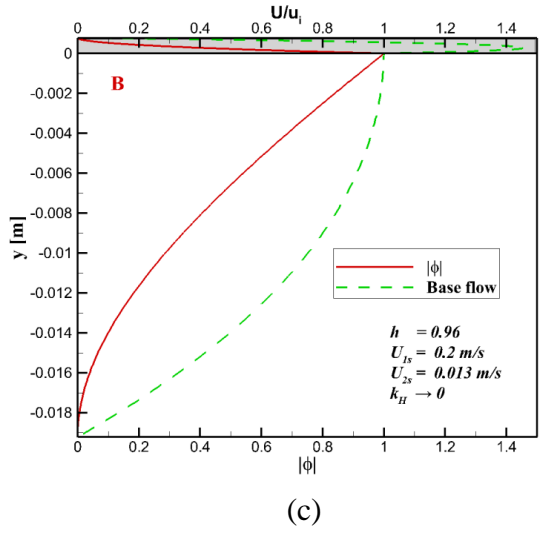
(c)

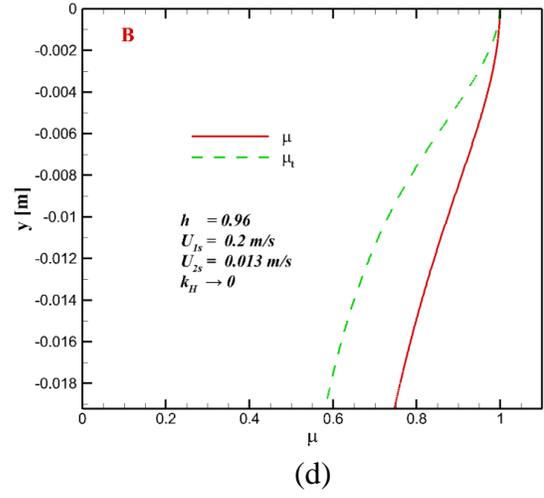
(d)

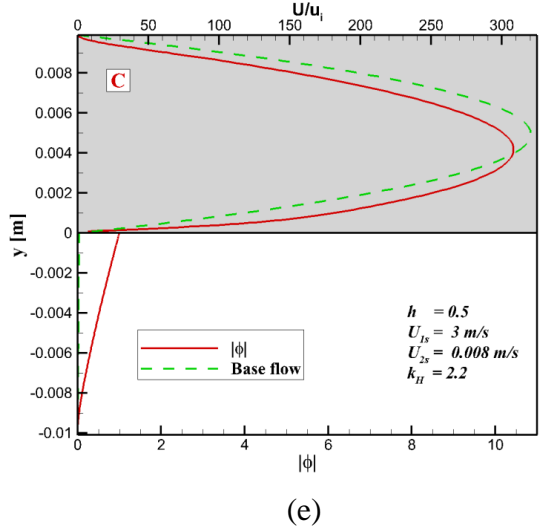
(e)

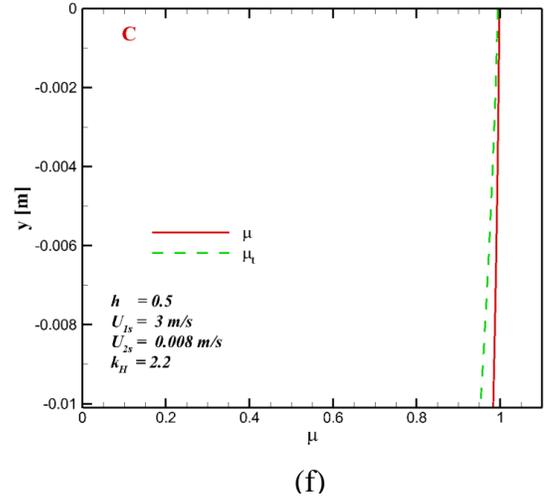
(f)

FIG. 3. (a), (c), (e) Amplitude of the stream function critical perturbation (eigenfunction, red solid line) and the base-flow velocity profile (green dashed lines). (b), (d), (f) Profiles of the dimensionless effective viscosity ($\mu$, red solid line) and the tangent viscosity ($\mu_t$, green dashed line) at points A, B, and C (see Fig. 2); y < 0 – CMC03, y > 0 (shaded region) – air.

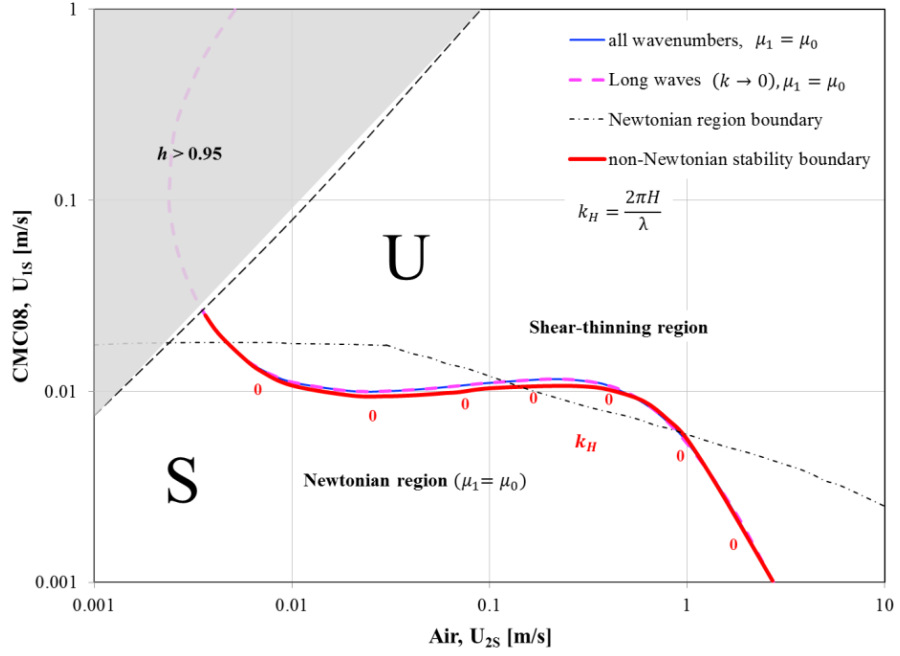

FIG. 4. Stability map for horizontal air-CMC08 flow $(r=1000; m=54071; \rho_2 =1\text{kg/m}^3;$ $\mu_2 =1.81\cdot 10^{-5}\text{Pa}\cdot\text{s}; \sigma=0.072\,\text{N/m})$ in a $H=0.02\text{m}$ channel. $S$ and $U$ denote the stable and unstable regions, respectively.

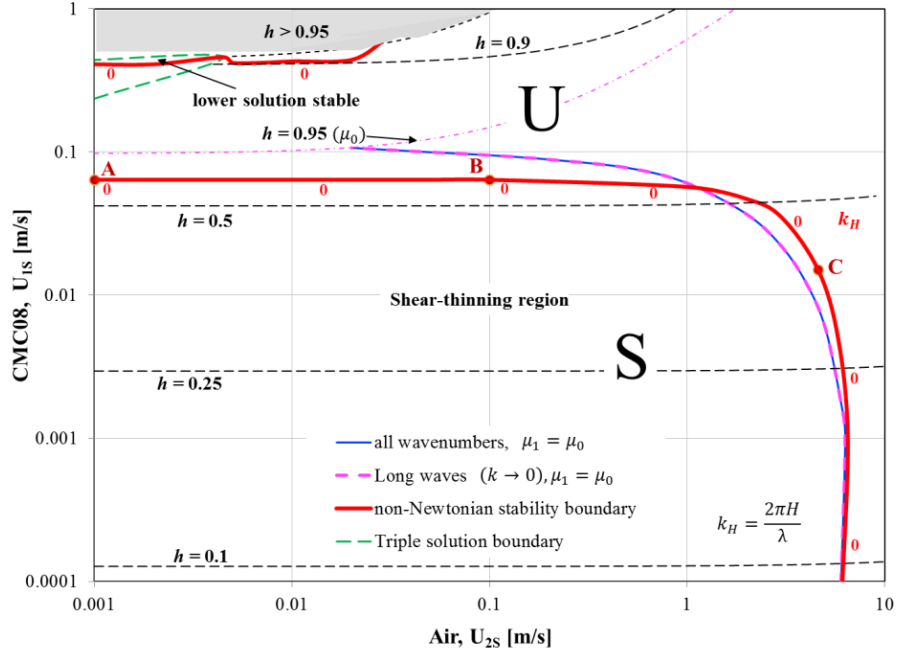

FIG. 5. Stability map for 5° downward inclined air-CMC08 flow $(r=1000; m=54071; \rho_2 =1\text{kg/m}^3;$ $\mu_2 =1.81\cdot 10^{-5}\text{Pa}\cdot\text{s}; \sigma=0.072\,\text{N/m})$ in a $H=0.02\text{m}$ channel. $S$ and $U$ denote the stable and unstable regions, respectively.

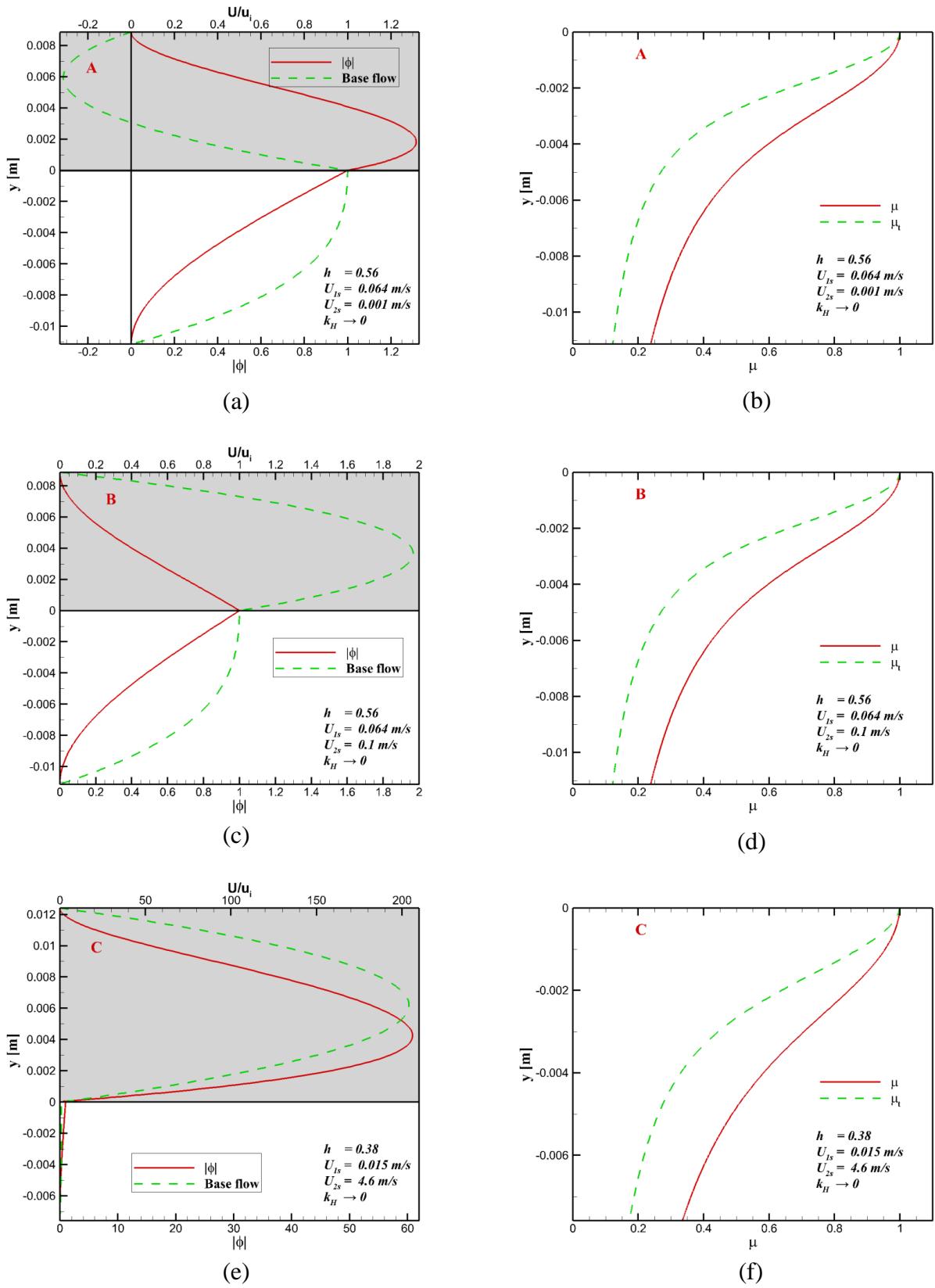

FIG. 6. (a), (c), (e) Amplitude of the stream function critical perturbation (eigenfunction, red solid line) and the base-flow velocity profile (green dashed line). (b), (d), (f) Profiles of the dimensionless effective viscosity ($\mu$, red solid line) and the tangent viscosity ($\mu_t$, green dashed line) for 5° downward inclined air-CMC08 flow at points A, B, and C (see Fig. 5); y < 0 – CMC08, y > 0 (shaded region) – air.

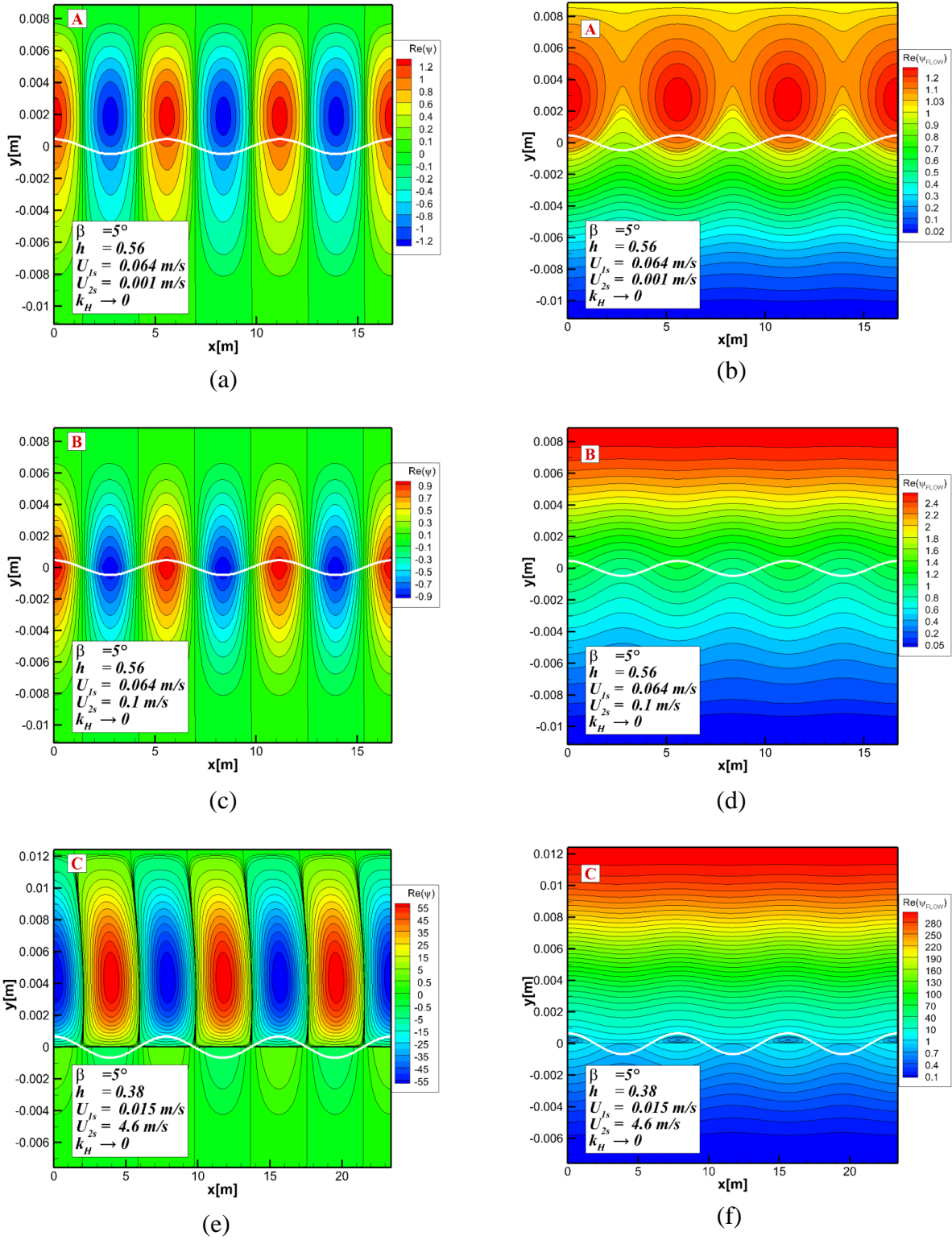

FIG. 7. Contours of the stream function critical perturbation ($\text{Re}(\psi)$, (a), (c), and (e)) and of the critically perturbed flow: base flow + perturbation ($\text{Re}(\psi_{\text{FLOW}})$, (b), (d), and (f)), for 5° downward inclined air-CMC08 flow at points A, B, and C (see Fig. 5, 6). White solid line is the corresponding disturbed interface. $x$ is the flow direction, $0 \leq x \leq 3l_{wave}$.

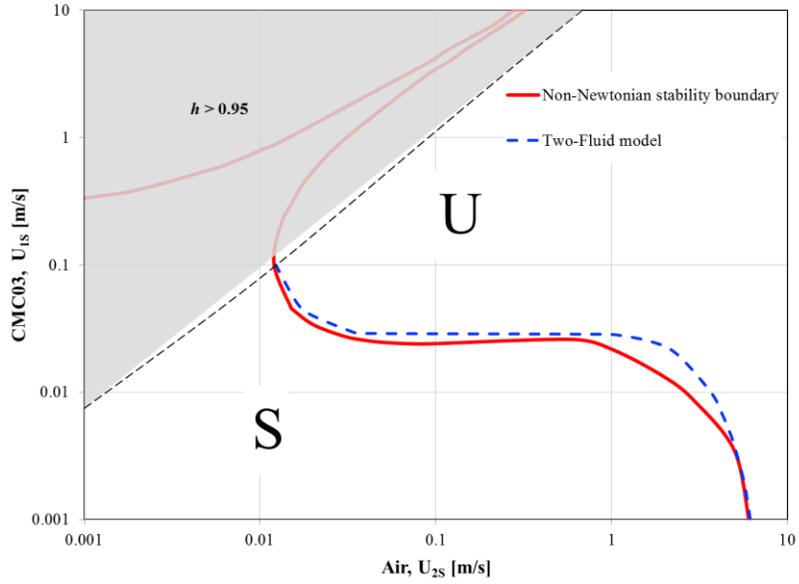

FIG. 8. Stability boundaries for horizontal air-CMC03 flow: Comparison of the exact analysis and the Two-Fluid model of Kushnir et al. (2017). *S* and *U* denote the stable and unstable regions, respectively.

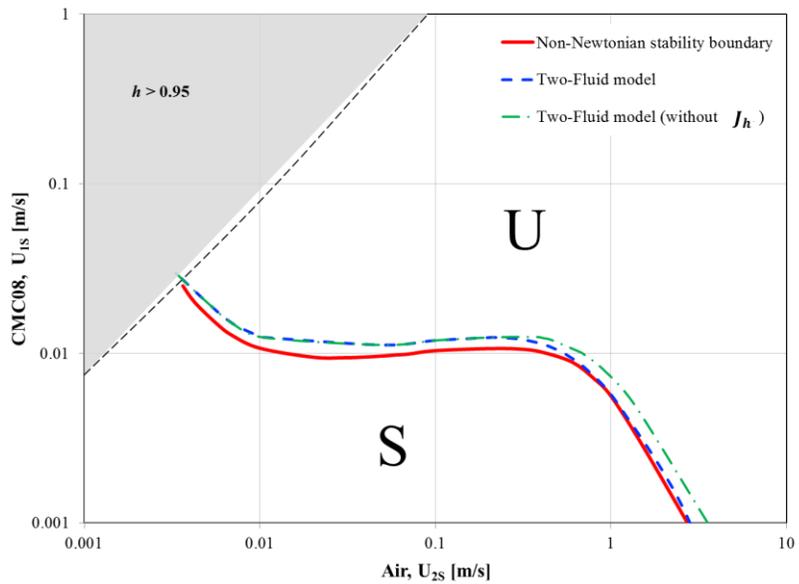

FIG. 9. Stability boundaries for horizontal air-CMC08 flow: Comparison of the exact analysis and the Two-Fluid model of Kushnir et al. (2017) with and without the 'sheltering' term, $J_h$. *S* and *U* denote the stable and unstable regions, respectively.

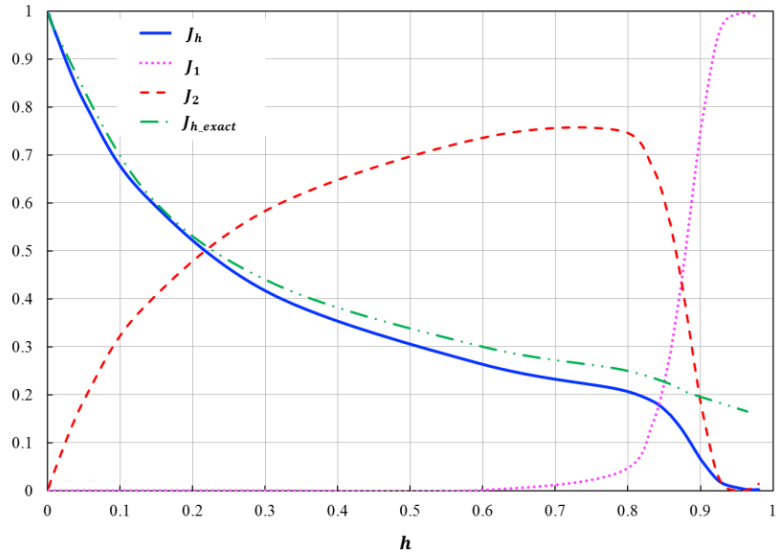

FIG. 10. Values of the *J* terms along the TF model stability boundary for horizontal air-CMC08 flow.

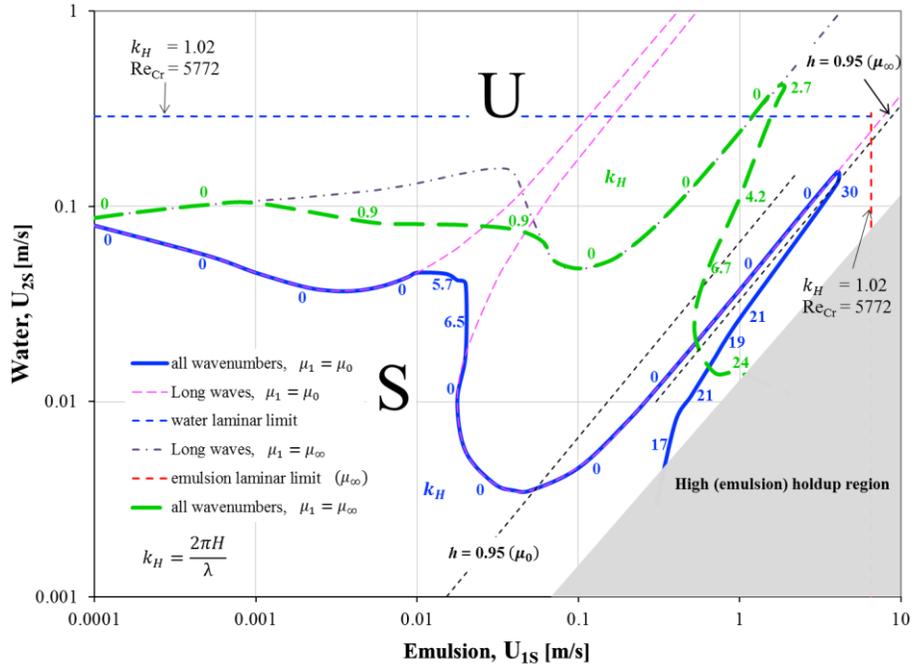

FIG. 11. Stability boundaries for two horizontal Newtonian liquid/water systems $(1/r = 1.25; \rho_2 = 1000 \text{kg/m}^3; \mu_2 = 0.001 \text{Pa} \cdot \text{s}; \sigma = 0.03 \text{N/m}, H = 0.02\text{m})$: Newtonian liquid viscosity – $\mu_0 = 0.723 \text{Pa} \cdot \text{s}$ (blue solid boundary), $\mu_\infty = 0.0181 \text{Pa} \cdot \text{s}$ (green dashed boundary). Note that *h* denotes the emulsion holdup. *S* and *U* denote the stable and unstable regions, respectively.

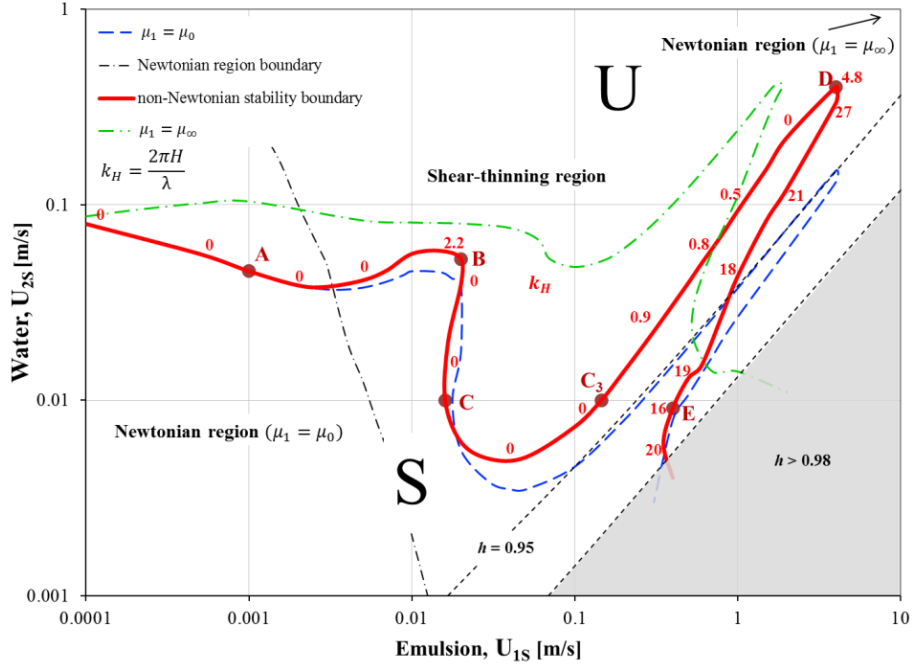

FIG. 12. Stability boundary for horizontal non-Newtonian oil-in-water emulsion/water flow $(m=723; m_\infty=18.1; r=0.8; \rho_2=1000\,\text{kg/m}^3; \mu_2=0.001\,\text{Pa}\cdot\text{s}; \sigma=0.03\,\text{N/m})$ in a $H=0.02$m channel in comparison with the $\mu_0$ and $\mu_\infty$-Newtonian stability boundaries. Note that $h$ denotes the emulsion holdup. $S$ and $U$ denote the stable and unstable regions, respectively.

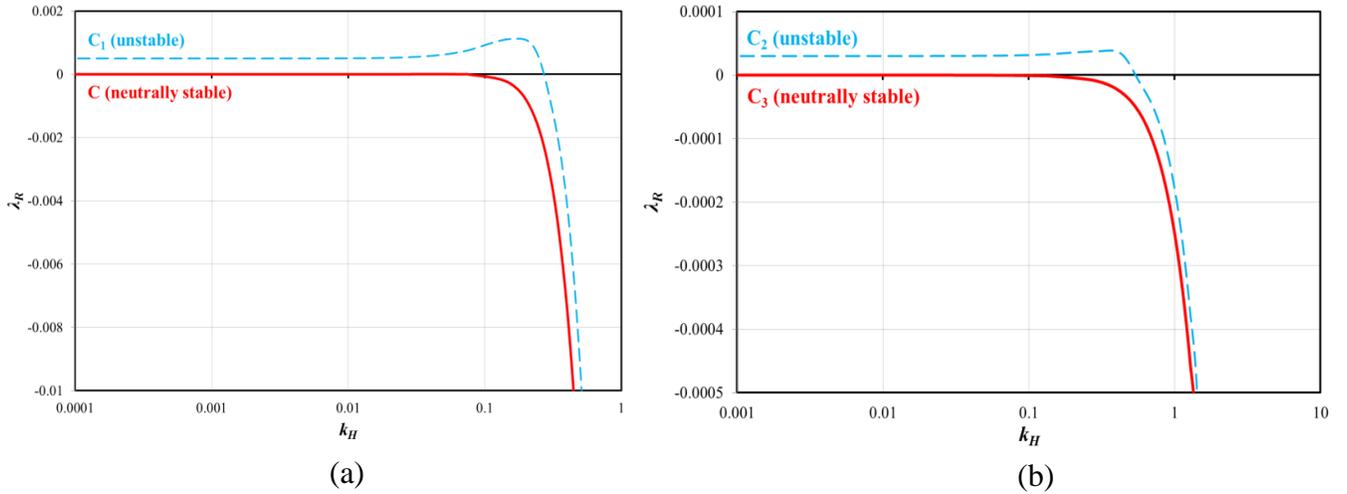

FIG. 13. Growth rate of perturbation vs. wavenumber. (a) at point C ($U_{1S}=0.016$ m/s; $U_{2S}=0.01$ m/s; $h=0.86$, red solid line) on the neutral stability curve (see Fig. 12) and in its vicinity in the unstable region, at point $C_1$ ($U_{1S}=0.017$ m/s; $U_{2S}=0.01$ m/s; $h=0.87$, light blue dashed line). (b) at point $C_3$ ($U_{1S}=0.147$ m/s; $U_{2S}=0.01$ m/s; $h=0.93$, red solid line) on the neutral stability curve (see Fig. 12) and in its vicinity in the unstable region, at point $C_2$ ($U_{1S}=0.146$ m/s; $U_{2S}=0.01$ m/s; $h=0.93$, light blue dashed line).

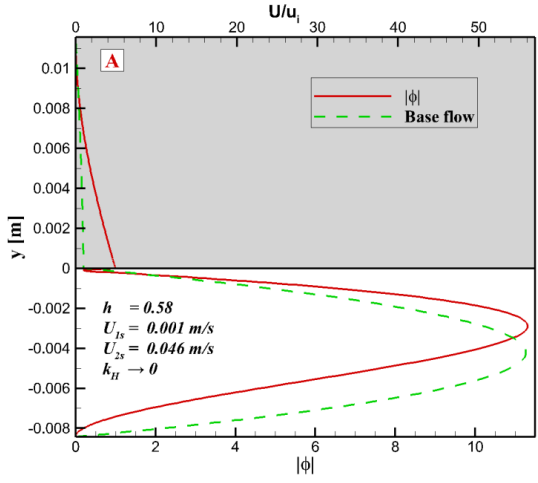

(a)

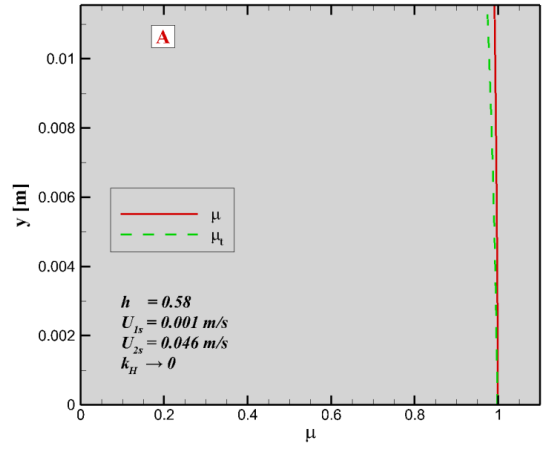

(b)

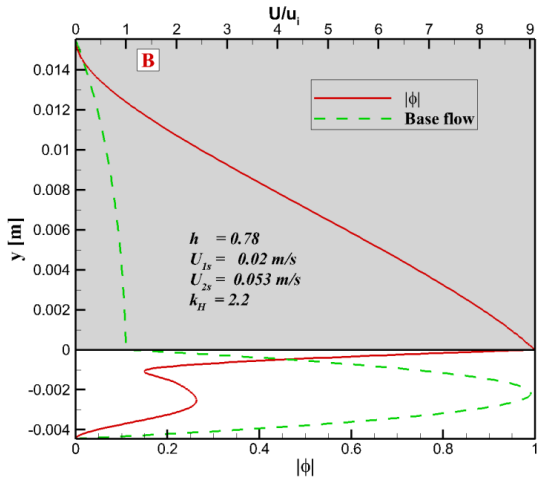

(c)

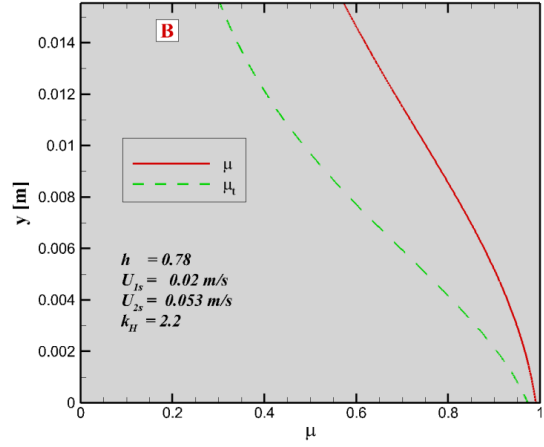

(d)

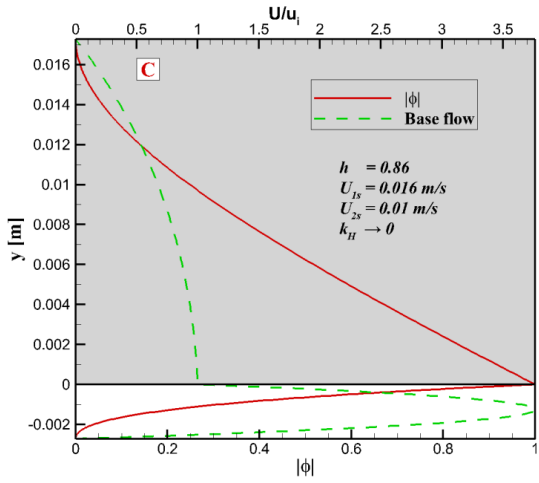

(e)

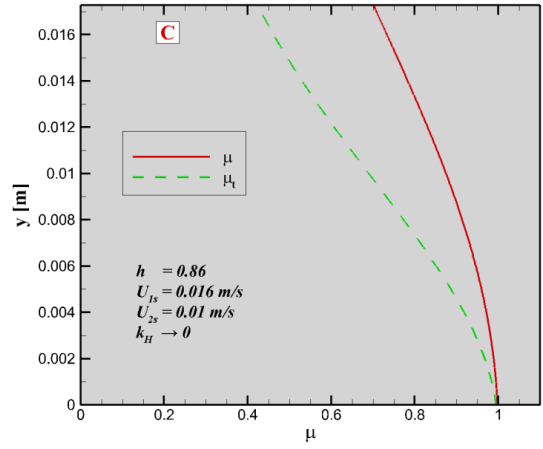

(f)
36

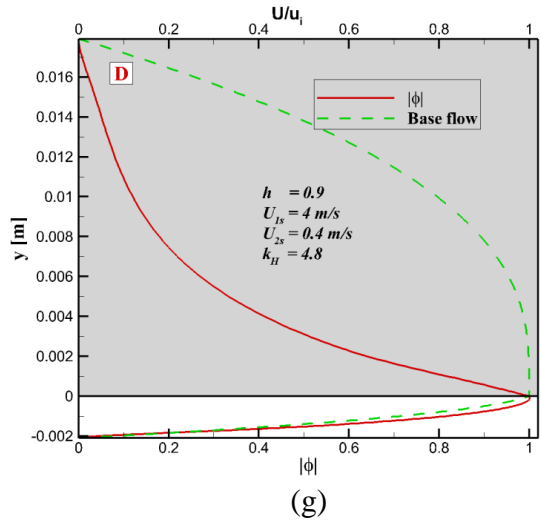

(g)

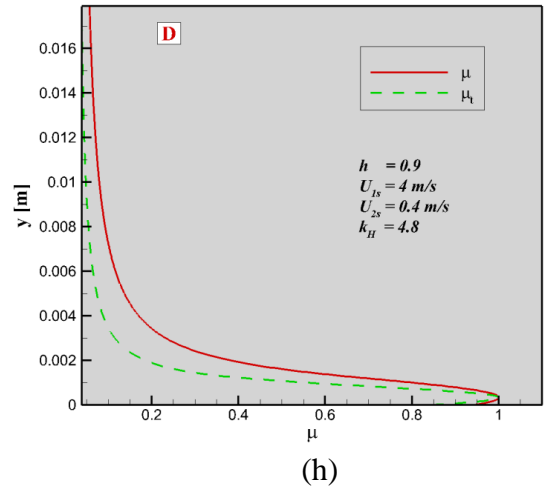

(h)

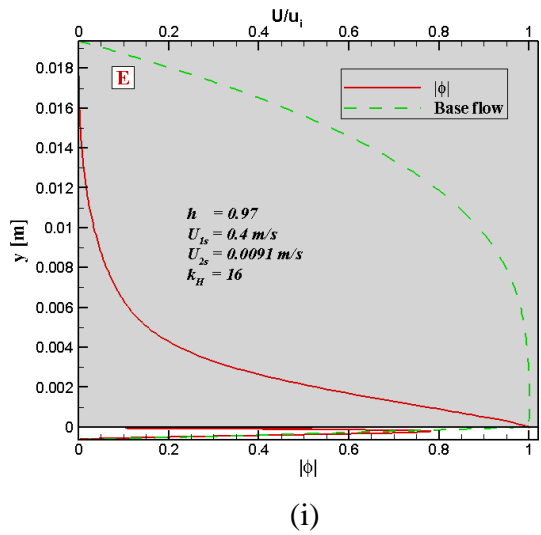

(i)

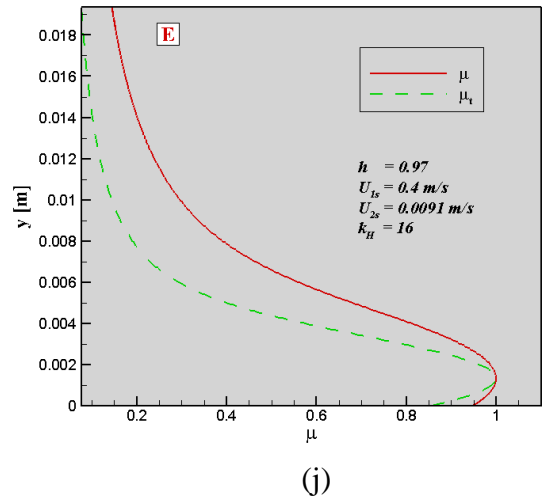

(j)

FIG. 14. (a), (c), (e), (g), (i) Amplitude of the stream function critical perturbation (eigenfunction, red solid line) and the base-flow velocity profile (green dashed line). (b), (d), (f), (h), (j) Profiles of the dimensionless effective viscosity ($\mu$, red solid line) and the tangent viscosity ($\mu_t$, green dashed line) for horizontal oil-in-water emulsion/water flow at points A, B, C, D and E (see Fig. 13); $y < 0$ – water, $y > 0$ (shaded region) – emulsion.